\documentclass[11pt]{article}
\usepackage{amssymb,amsmath}
\usepackage{bm} %math bold symbols
\usepackage{booktabs} %some alignment stuff
\usepackage{array}
\usepackage{latexsym}
\usepackage{graphicx}
\usepackage{color}
\usepackage{datetime}
\usepackage[nosort]{cite}
\usepackage{verbatim}
\usepackage{enumerate}
\usepackage{enumitem}
\usepackage{chngpage} % allows for temporary adjustment of side margins
\usepackage{mathrsfs}
\usepackage{euscript}
\usepackage{psfrag}

\usepackage{mciteplus}

\usepackage[colorlinks=true,      linkcolor=blue,      urlcolor=blue,
            filecolor=blue,      citecolor=blue,       pdfstartview=FitH,
                        pdfpagemode=UseNone,      bookmarksopen=true]{hyperref}  % LINKS
\usepackage[all]{hypcap}     %should be loaded AFTER hyperref, which should otherwise be loaded last.

\usepackage{tensor}
\usepackage{physics}
\usepackage{tikz}
\usepackage{mathtools}

\renewcommand{\comm}[1]{} %for commenting out blocks of text

\DeclareMathOperator*{\diag}{diag}
\DeclareMathOperator*{\arctanh}{arctanh}

\def\({\left(}
\def\){\right)}
\def\[{\left[}
\def\]{\right]}

\def\coeff#1#2{{\textstyle \frac{#1}{#2}}}

\def\One{{\hbox{ 1\kern-.8mm l}}}

\def\barray{\begin{array}}
\def\earray{\end{array}}
\def\be{\begin{equation}}
\def\ee{\end{equation}}
\def\bea{\begin{eqnarray}}
\def\eea{\end{eqnarray}}
\def\bal{\begin{align}}
\def\eal{\end{align}}

\numberwithin{equation}{section} % replaces the hack below

\definecolor{cardinal}{rgb}{0.6,0,0}
\definecolor{darkgreen}{rgb}{0,0.4,0}
\definecolor{golden}{rgb}{0.92, 0.7, 0}
\definecolor{midnight}{rgb}{0, 0, 0.5}
\definecolor{darkblue}{rgb}{0, 0, 0.7}
\definecolor{darkred}{rgb}{0.6, 0, 0}
\definecolor{purple}{rgb}{0.5, 0, 0.5}

\def\oneone{\rlap 1\mkern4mu{\rm l}}

\def\Neql#1{{\cal N}\!=\!{#1}}

\def\IR{\mathbb{R}}

\def\IT{\mathbb{T}}

\def\ZZ{\mathbb{Z}}

\def\cD{{\cal D}}
\def\cF{{\cal F}}

\def\cI{{\cal I}}

\def\cL{{\cal L}}
\def\cM{{\cal M}}
\def\cN{{\cal N}}
\def\cO{{\cal O}}
\def\cP{{\cal P}}
\def\cQ{{\cal Q}}
\def\cR{{\cal R}}

\def\cX{{\cal X}}

\def\scrC{{ \mathscr{C}}}

\def\RAdS{{R_{AdS}}}
\def\RAdSsq{{R^2_{AdS}}}

\def\muscal{{\lambda}}
\def\newmuscal{{\mu}}
\begin{document}

\phantom{AAA}
\vspace{-10mm}

\begin{flushright}
%
%IPHT-T19/021\\
%
\end{flushright}

\vspace{1.9cm}

\begin{center}

{\huge {\bf New Superstrata  from Three-Dimensional Supergravity }}\\
{\huge {\bf \vspace*{.25cm}  }}

\vspace{1cm}

{\large{\bf { Bogdan Ganchev$^{1}$, Anthony Houppe$^{1}$   and  Nicholas P. Warner$^{1,2,3}$}}}

\vspace{1cm}

$^1$Institut de Physique Th\'eorique, \\
Universit\'e Paris Saclay, CEA, CNRS,\\
Orme des Merisiers, Gif sur Yvette, 91191 CEDEX, France \\[12 pt]
\centerline{$^2$Department of Physics and Astronomy}
\centerline{and $^3$Department of Mathematics,}
\centerline{University of Southern California,} 
\centerline{Los Angeles, CA 90089, USA}

\vspace{10mm} 
{\footnotesize\upshape\ttfamily anthony.houppe @ ipht.fr, warner @ usc.edu} \\

\vspace{2.2cm}
 
\textsc{Abstract}

\end{center}

\begin{adjustwidth}{3mm}{3mm} % to adjust the L and R margins
 
\vspace{-1.2mm}
\noindent
We find a  two-parameter family of {\it generalized superstrata} that emerge as  smooth, supersymmetric solutions in three-dimensional gauged supergravity coupled to additional scalar fields.   This new family of generalized superstrata are smooth microstate geometries and may be thought of as supersymmetric Coulomb-branch extensions of the original superstrata in which  the underlying supertube undergoes an elliptical deformation. These solutions had already been obtained numerically, and as series solutions, to the equations of motion, and some of them were conjectured to be supersymmetric.  Here we prove the supersymmetry of an entire two-parameter family and we obtain a highly non-trivial analytic and smooth solution for a one-parameter limit in which the global symmetry of the metric is enhanced to $SO(3)$.  We also confirm that the other known families of microstrata are {\it not} supersymmetric. We conclude with a cursory analysis of some of the singular brane distributions that can be accessed from three-dimensional gauged supergravity while preserving the same supersymmetries as the superstratum, and therefore of the three-charge black hole. 

\end{adjustwidth}

%\end{titlepage}
\thispagestyle{empty}
\newpage

%%%%%%%%%%%%%%%%%%%%%%%%%%%%%%%%%%%%%

\baselineskip=17pt
\parskip=5pt

\setcounter{tocdepth}{2}
\tableofcontents
%\newpage

\baselineskip=15pt
\parskip=3pt

\newpage

%%%%%%%%%%%%%%%%%%%%%%%%%%%%%%%%%%%%%
\section{Introduction}
\label{sec:Intro}
%%%%%%%%%%%%%%%%%%%%%%%%%%%%%%%%%%%%%

The  D1-D5-P system has proven to be one of the most powerful routes to understanding black-hole microstructure. In no small way, this power is derived from the fact that the states of the system can be captured by a CFT on the common spatial direction of the D1 and D5 branes.  These branes back-react to create a geometry, AdS$_3$ $\times S^3$  $\times \cM_4$, where $\cM_4$ is either $K3$ or $\IT^4$.  Superstrata are then the smooth geometries that result from back-reacting coherent combinations of particular left-moving momentum excitations in the CFT  \cite{Bena:2015bea,Bena:2016ypk,Tian:2016ucg,Bena:2016agb,Bena:2017xbt,Heidmann:2019zws,Heidmann:2019xrd,Shigemori:2020yuo}.   The holographic dictionary for these excitations and the dual geometries has been thoroughly examined and has passed numerous,  precise, non-trivial tests \cite{Kanitscheider:2006zf,Kanitscheider:2007wq,Giusto:2004id,Ford:2006yb, Lunin:2012gp, Giusto:2013bda,Bena:2015bea,Bena:2016ypk,Bena:2017xbt,Bakhshaei:2018vux,Rawash:2021pik}.   We thus have very high confidence in the role of superstrata in describing particular sectors of the BPS black-hole microstructure.
 
One of the first steps in the construction of superstrata is to reduce from IIB string theory to $(1,0)$ supergravity coupled to tensor multiplets in six dimensions.  The compactification on $\IT^4$ is elementary and the reduced six-dimensional supergravity system emerges from the choice of the CFT states one is seeking to describe \cite{Bena:2015bea}.   A significant sector of the six-dimensional BPS equations have the virtue of being linear \cite{Bena:2011dd} and this has led to a huge literature on the construction and study of many families of superstrata.  

Despite this progress, the six-dimensional equations of motion, and even the BPS equations, are very non-trivial.  Indeed, the known superstrata are based on some major simplifying assumptions to reduce the  complexity of the BPS system.  Thus the  classification of superstratum solutions is far from complete.  More generally, the construction of ``microstrata,''  non-supersymmetric microstate geometries dual to non-BPS excitations of the D1-D5 CFT, is wide-open territory.

As always, the art of analytic construction of  new supergravity backgrounds is to find a way to reduce the degrees of freedom so as to render the problem solvable, while retaining enough degrees of freedom to make the solution new and interesting.  The gauged supergravities that emerge from  dimensional reduction on spheres have frequently proven to be an intriguing way to achieve this goal. The power of the technique lies in the fact that one can obtain exact solutions to the higher dimensional supergravity theory by solving equations of motion (or BPS equations) in lower dimensions.  The limitation is that the gauged supergravity only describes a very limited set of degrees of freedom in the higher-dimensional theory.  Nevertheless, holographic field theory has many triumphs that came from lower-dimensional gauged supergravity, and it is hard to imagine how some of these physically interesting backgrounds could have been constructed without the power and insight gained from gauged supergravity (see, for example, \cite{deWit:1984nz,Freedman:1999gp,Pilch:2000ej,Pilch:2000ue,Pilch:2000fu,Warner:2000qh, Ahn:2000mf,Gubser:2000nd, Corrado:2001nv,Ahn:2001by,Ahn:2001kw,Ahn:2002qga,Pope:2003jp,Halmagyi:2005pn,Gowdigere:2005wq,Bobev:2010ib,Bobev:2011rv,Bobev:2013yra,Pilch:2015vha, Bobev:2018eer, Bena:2018vtu,Arav:2021tpk}).

A central part of establishing a consistent truncation is to provide the ``uplift formulae:'' expressions that show precisely how the higher-dimensional fields can be reconstructed from their lower-dimensional counterparts. This provides the exact mapping through which solutions of the higher dimensional equations are obtained from the lower-dimensional solutions.  It has taken many years to construct some of these uplift formulae, but this process is now relatively well-understood, although it is still a very technical enterprise. 

It has long been known that there are consistent truncations of the D1-D5 system to some classes of gauged supergravity in three dimensions, via the AdS$_3$ $\times S^3$ vacuum background \cite{Cvetic:2000dm,Cvetic:2000zu, Nicolai:2001ac, Nicolai:2003bp, Nicolai:2003ux,Deger:2014ofa,Samtleben:2019zrh}.  However, in order for the consistent truncation to capture superstrata, one needs to extend the standard consistent truncation to include some extra six-dimensional tensor multiplets.  This consistent truncation, and its  uplift formulae, were     obtained and proven in \cite{Mayerson:2020tcl}. The corresponding supersymmetry transformations and BPS equations were then obtained in \cite{Houppe:2020oqp}.  This three-dimensional gauged supergravity contains the degrees of freedom necessary to describe the $(1,m,n)$-superstrata\footnote{For the meaning of this notation, see any of  \cite{Bena:2015bea,Bena:2016ypk,Bena:2017xbt,Heidmann:2019zws,Heidmann:2019xrd,Shigemori:2020yuo,Mayerson:2020tcl}.}.

While the construction  of new consistent truncations and three-dimensional gauged supergravities is interesting in its own right, the broader goal of the work in \cite{Mayerson:2020tcl, Houppe:2020oqp} was to obtain a system that contains superstrata and was sufficiently simple (because of its low dimensionality)  that one might find non-supersymmetric ``microstrata.''   A very important first step along these lines was achieved in  \cite{Ganchev:2021pgs}, in which multiple families of non-extremal microstrata were constructed perturbatively and numerically.

 One of the surprises in   \cite{Ganchev:2021pgs} was the appearance of what seemed to be a hithertofore unknown branch of superstrata.  That is, there appeared to be a new branch of {\it generalized superstrata} in which most of the fields only differ  slightly from those of the standard superstratum, except for one scalar which vanishes for the latter, but is turned on for the new solutions.  This scalar controls an elliptical deformation of the underlying supertube, and, as we will discuss, appears to be exploring a Coulomb branch of the original superstratum. There are thus two parameters in generalized superstrata: one is the original superstratum amplitude, and the other controls the elliptical deformation.  As one might expect from such a deformation, the geometry of a generalized superstratum  has less symmetry than the original superstratum.

 Indirect evidence for the supersymmetry of this branch was given in the construction of microstrata \cite{Ganchev:2021pgs}, although only when one of the two defining parameters of the solution took on a special value. Using high orders of perturbation theory it was shown that for a particular family of microstrata, the microsratum frequency of the excitations was unshifted from its supersymmetric value and that the central IR core of the solution limited to the supersymmetric ground state. One of the purposes of this paper is to show that this branch of solutions, found in  \cite{Ganchev:2021pgs},  is indeed fully supersymmetric and this is true \textit{for all values} of the parameters. As part of establishing this supersymmetric family, we also find that all the other microstrata discovered in  \cite{Ganchev:2021pgs} break all the supersymmetry.
  
We will also investigate another aspect of what are known as supersymmetric Coulomb branch flows.   Such flows usually involve singular infra-red limits in which the underlying branes spread into a distribution, or in which new brane-sources are introduced.  One can obviously spread the D1 and D5 branes out into harmonic distributions but one should remember that the global AdS$_3$ $\times S^3$ of the D1-D5 system is based on a supertube with its attendant angular momentum and KKM dipole distribution.  We will use the three-dimensional supergravity  to investigate the Coulomb branch excitations of this system.  As one would expect from the corresponding results in five dimensions \cite{Freedman:1999gk,Khavaev:2001yg, Gowdigere:2005wq}, the consistent truncation leads to very limited families of singular brane distributions in the IR.   

An interesting facet of these Coulomb branch flows is the supersymmetry that they preserve. It is natural to define the supersymmetry of a Coulomb branch flow through the projectors that are obtained from the product of the gamma matrices that run parallel to the brane configuration.  This is the choice made in many analyses in five dimensions (see, for example, \cite{Freedman:1999gp, Freedman:1999gk,Khavaev:2001yg, Gowdigere:2005wq}).  A similar choice was analyzed in \cite{Deger:2019jtl}  for Coulomb branch flows in three dimensions, and it resulted in domain-wall solutions with singular brane distributions and sometimes rather singular asymptotics at infinity.  In this paper we will make a different choice for the supersymmetry, one that is motivated by the superstrata and black-hole microstructure, and we will find some rather different solutions with singular brane sources.

We will need to draw upon several sources to set up our computations.  First, we will need the details of the three-dimensional supergravity and its supersymmetry transformations. We will therefore summarize the results of  \cite{Houppe:2020oqp} in Section \ref{sec:3D-Sugr}.  In Section \ref{sec:truncations} we will summarize the reduced system of equations and degrees of freedom used in  \cite{Ganchev:2021pgs} that lead to the new branch of superstrata.   In Section \ref{sec:preparing_computation}  we assemble all the pieces that will be needed to obtain the BPS equations  and in Section \ref{sec:the_susy_equations}  we put it all together,  obtaining the BPS equations. Section  \ref{sec:summary_of_the_equations} contains a summary of the system.  In this paper, we will not make an exhaustive analysis of all solutions to the BPS system, but we will reduce the system so that we can analyze the supersymmetry of microstrata, and exhibit a more standard supersymmetric Coulomb branch flow. 

 We  specialize the BPS equations to microstrata in Section \ref{sec:generalized_superstrata}   and obtain the first order system  that governs {\it generalized superstrata}.   We find that these new families of superstrata involve seven unknown functions of one variable:  there are five non-linear, first-order BPS equations, which must be supplemented by two of the equations of motion.  As we will discuss in Section \ref{sec:Conclusions}, we believe that the need for two of the equations of motion  reflects the fact that the generalized superstrata have a two-dimensional Coulomb branch.  We further specialize this   system to  a  {\it simple branch} identified in   \cite{Ganchev:2021pgs}.  This branch is governed by four unknown functions that are determined entirely by four  first-order BPS equations.   We exhibit a complete, smooth analytic family of solutions to this system of equations.     In Section \ref{sec:the_coulomb_branch} and Appendix \ref{app:coulomb_solutions}  we examine a more standard, and more symmetric family of pure Coulomb branch flows,  and we uplift these solutions to six dimensions.   We leave the uplift of the new superstrata to future work as the geometry has very little symmetry and preliminary computations reveal a startling level of complication.  We will discuss this, and make further comments about our work in Section \ref{sec:Conclusions}.

Our notation and conventions will be  those of   \cite{Houppe:2020oqp} and we summarize them in Appendix \ref{app:Conventions}.  

%%%%%%%%%%%%%%%%%%%%%%%%%%%%%%%%%%%%%
\section{The three-dimensional supergravity}
\label{sec:3D-Sugr}
%%%%%%%%%%%%%%%%%%%%%%%%%%%%%%%%%%%%%
%\vspace{0.8cm}

%%%%%%%%%%%%%%%%%%%
\subsection{The fields and their symmetries}
\label{sec:fields}
%%%%%%%%%%%%%%%%%%%

The $(1,m,n)$ superstrata can be described within an $\Neql{4}$ gauged supergravity theory in three dimensions.  This theory has a graviton, four gravitini, $\psi_\mu^A$, two sets of six vector fields, $A_\mu^{IJ}= -A_\mu^{JI}$ and $B_\mu^{IJ}= -B_\mu^{JI}$, $20$ fermions, $\chi^{\dot A r}$, and  $20$ scalars parametrized by the coset\footnote{See Appendix \ref{app:Conventions} fo a more complete discussion of the group theory and our conventions. 
}:
\begin{equation}
\frac{G}{H} ~\equiv~ \frac{SO(4,5)}{SO(4) \times SO(5)} \,.
\label{coset1}
\end{equation}
The scalar fields are parametrized by a coset representative, $L(x)$, that is viewed as transforming on the left under a global action of $g \in G$, and on the right under a composite, local symmetry $h(x) \in H$:  $L(x) \to g L(x) h (x)^{-1}$.  The gauge group is an $SO(4) \ltimes \IT^6$ subgroup of $G$ and the gauging promotes  this  to a local  action on $L(x)$: 
\begin{equation}
L(x)\; \longrightarrow \; g_0 (x)\, L(x) \,h^{-1}(x)\;, \qquad
  g_0 (x) \in SO(4) \ltimes \IT^6  \; , \; h(x) \in H \,.
\label{GHaction}
\end{equation}

The $\IT^6$ gauge symmetry is typically fixed by setting six of the scalars to zero, and the associated gauge fields, $B_\mu^{IJ}$, can be integrated out.  The result is an action for the graviton, the gravitini, the fermions, the  $SO(4)$ gauge fields, $A_\mu^{IJ}$, and $14$ scalars.  The scalars can be described by the non-compact generators of a  $GL(4,\IR)$ matrix, ${P_I}^J$, and an $SO(4)$ vector,  $\chi_I$.  The theory then has a remaining gauge symmetry $SO(4) \subset G$ that acts on the $\chi_I$ and on the left of  ${P_I}^J$, along with   a composite, local symmetry $SO(4) \subset H$ that acts on the fermions and on the right of ${P_I}^J$. This part of the composite local symmetry is typically fixed by taking $P$ to be symmetric.

Following \cite{Mayerson:2020tcl} and \cite{Houppe:2020oqp} it is easiest to characterize the fermionic sector using the structure that it inherits by embedding it in a larger gauged $\Neql{8}$ supergravity in three dimensions.  This is simply a convenient mathematical trick  to enable us to use the  real representation of the $SO(8)$ gamma matrices.  Thus we will use gamma matrices, $\Gamma^{I}_{A\, \dot{A}}$ where all indices ( $I, A , \dot{A}$) run over $1, \dots,8$, but $I, J \dots$ represent the vector of $SO(8)$ and $A , \dot{A},B , \dot{B}, \dots$ are the spinor indices of  $8^+$ and $8^-$.   One should note that it is only on the gamma matrices that one has $I, J = 1, \dots,8$.  The $\Neql{4}$ gauged supergravity  requires the restriction of  $I, J = 1, \dots,4$ and so everywhere other than on the gamma matrices we take $I, J = 1, \dots,4$, and view these indices as transforming under the vector of $SO(4)$.  In particular, the gauge fields $A_\mu^{IJ}= -A_\mu^{JI}$ transform, of course, as the adjoint of $SO(4)$.

The gravitini, $\psi_\mu^A$, the fermions, $\chi^{\dot A r}$ and the supersymmetries $\epsilon^A$ {\it a priori} transform as $SO(8)$ spinors, but they are reduced by half through the projection conditions:
\begin{equation}
\big(\oneone  ~-~ \Gamma^{5678}\big) \,  \Phi ~=~ 0\,,
\label{SO8proj1}
\end{equation}
where $\Phi$ is any fermion, including the supersymmetries.  Because the $SO(8)$ helicity projector is $\Gamma^{12345678}$, this condition translates to:
\begin{equation}
\big(\oneone  ~-~ \Gamma^{1234}\big)_{AB} \, \epsilon^B  ~=~ 0 \,, \qquad \big(\oneone  ~-~ \Gamma^{1234}\big)_{AB} \,\psi^B_\mu  ~=~ 0 \,, \qquad \big(\oneone  ~+~ \Gamma^{1234}\big)_{\dot A \dot B}\, \chi^{\dot B r}  ~=~ 0   \,.
\label{SO8proj2}
\end{equation}
This then defines the fermionic sector of the $\Neql{4}$ theory.

One can also characterize the $\Neql{4}$  subsector of the $\Neql{8}$ theory  via a group invariance.   The group $SO(8)$ contains  $(SU(2))^4$ and one of the $SU(2)$ rotations makes self-dual rotations on the indices $5,6,7,8$.  The $\Neql4$ theory is the singlet sector of this $SU(2)$.

%%%%%%%%%%%%%%%%%%%
\subsection{The gauge fields and scalar composites}
\label{sec:gauge}
%%%%%%%%%%%%%%%%%%%

The gauge covariant derivatives are defined in terms of the $SO(4)$-dual vectors:
\begin{equation}
{\widetilde A_\mu}{}^{IJ}  ~\equiv~ \coeff{1}{2} \,\epsilon_{IJKL}\,{A_\mu}^{KL} \,,  
\label{dualGFs}
\end{equation}
with minimal couplings 
\begin{equation}
\cD_\mu \, \cX_{I}  ~=~ \partial_\mu \, \cX_{I}   ~-~  2\, g_0\,\widetilde A_\mu{}^{IJ} \, \cX_{J} \,. 
\label{covderiv}
\end{equation}
The field strengths are thus
\begin{equation}
F_{\mu \nu}{}^{IJ}  ~=~ \coeff{1}{2}\, \epsilon_{IJKL} \,  \widetilde F_{\mu \nu}{}^{KL}    ~=~  \partial_{\mu}   A_{\nu}{}^{IJ}   ~-~  \partial_{\nu}   A_{\mu}{}^{IJ}   ~-~ 2 \,  g_0 \, \big(   A_{\mu} {}^{IL} \,\widetilde  A_{\nu} {}^{LJ}  ~-~ A_{\mu} {}^{JL} \,\widetilde  A_{\nu} {}^{LI}\big)  \,.
\label{fieldstrength}
\end{equation}
The gauge coupling has dimensions of inverse length, and is related to the charges of the D1-D5 system via: 
 \begin{equation}
g_0 ~\equiv~ (Q_1 Q_5)^{-\frac{1}{4}} \,.
\label{g0reln}
\end{equation}

Because the gauge action only acts on the left of $P$, its covariant derivative is:
\begin{equation}
 \big(\cD_\mu \, P\big)_{I}{ }^{\, J}   ~=~ \partial_\mu \, P_{I}{ }^{\, J}   ~-~  2\, g_0\,\widetilde A_\mu{}^{IK} \, P_{K}{ }^{\, J} \,,
 \end{equation}
It is convenient to introduce 
\begin{equation}
m_{IJ}   ~\equiv~   \big(P  \, P^T\big)_{IJ} \,, \qquad m^{IJ}   ~=~  \big ( (P^{-1})^T\,P^{-1}  \big)^{IJ}  \,.
\label{Mdefn}
\end{equation}
and its covariant derivative is given by:
\begin{equation}
\cD_\mu m_{IJ}   ~=~   \partial_\mu m_{IJ}  ~-~   2\, g_0\,\widetilde A_\mu{}^{IK} m_{KJ }~-~   2\, g_0\,\widetilde A_\mu{}^{JK} m_{IK }   \,   \,.
\label{Dmform}
\end{equation}

We will also define the following combinations of fields:
\begin{equation}
\begin{aligned}
Y_{\mu \, IJ}  ~\equiv~&  \chi_J \,\cD_\mu \chi_I ~-~  \chi_I \,\cD_\mu\chi_J \,, \qquad  \qquad   C_{\mu}^{IJ}  ~\equiv~  g_{\mu \rho} \,  \varepsilon^{\rho \sigma \nu} \, m_{IK} m_{JL}\,F_{\sigma \nu}^{KL}  \,,   \\
\scrC_{\mu}^{IJ}   ~\equiv~ & P^{-1}{}_{I}{}^K \,  P^{-1}{}_{J}{}^L \, C_\mu^{KL}~=~  g_{\mu \rho}  \, \varepsilon^{\rho \sigma \nu} \,    P{}_{I}{}^K \,  P{}_{J}{}^L \,F_{\sigma \nu}^{KL}\,.
\end{aligned}
\label{cs_definition}
\end{equation}
Our  conventions for $\varepsilon$ are discussed in Appendix \ref{sec:st-conventions}.

The components of the scalar kinetic terms and the composite local connection are then given by  \cite{Houppe:2020oqp}:
\begin{equation}
\begin{aligned}
\cQ_\mu^{IJ}   ~=~ & \coeff{1}{2}\, \Big[ \, \big(P^{-1} \cD_\mu \, P\big)_{I}{ }^{\, J}  ~-~  \big(P^{-1} \cD_\mu \, P\big)_{J}{}^{\, I} ~+~\scrC_{\mu}^{IJ} \   \,  \Big] \,, \\
\cP_\mu^{Ir}   ~=~ & \coeff{1}{2}\, \Big[ \, \big(P^{-1} \cD_\mu \, P\big)_{I}{ }^{\, r}  ~+~  \big(P^{-1} \cD_\mu \, P\big)_{r}{}^{\, I} ~-~\scrC_{\mu}^{Ir} \   \,  \Big] \,, \qquad 1 \le r  \le 4 \,, \\
\cQ_\mu^{rs}   ~=~ & - \coeff{1}{2}\, \Big[ \, \big(P^{-1} \cD_\mu \, P\big)_{r}{ }^{\, s}  ~-~  \big(P^{-1} \cD_\mu \, P\big)_{s}{}^{\, r} ~+~\scrC_{\mu}^{rs} \   \,  \Big] \,,  \qquad 1 \le r,s  \le 4 \,,\\
\cQ_\mu^{r5}   ~=~&   \coeff{1}{\sqrt{2}}\, \big(P^{-1} \big)_{r}{ }^{\, J}  \, \cD_\mu \chi_J  \,,  \qquad
\cP_\mu^{I5}   ~=~  - \coeff{1}{\sqrt{2}}\, \big(P^{-1} \big)_{I}{ }^{\, J}  \,\cD_\mu  \chi_J \,,  \qquad 1 \le r  \le 4  \,.
 \end{aligned}
\label{ScalarDers}
\end{equation}
%

%%%%%%%%%%%%%%%%%%%
\subsection{The action}
\label{sec:3Daction}
%%%%%%%%%%%%%%%%%%%

Following \cite{Houppe:2020oqp},  we will use a metric signature of $(+- -)$.    The action may now be written as  \cite{Mayerson:2020tcl,Houppe:2020oqp} (note that the first reference uses different metric conventions): 
\begin{equation}
\begin{aligned}
\cL ~=~ & -\coeff{1}{4} \,e\,R ~-~  \coeff{1}{2}\,i e \, \overline{\chi}^{\dot{A} r}\gamma^\mu D_\mu\chi^{\dot{A} r} ~+~ \coeff{1}{2}  \,\epsilon^{\mu\nu\rho} \, \overline{\psi}^A_\mu D_\nu\psi^A_\rho   
 ~+~ \coeff{1}{8}\,e \, g^{\mu \nu} \, m^{IJ}   \, (\cD_\mu\, \chi_{I})  \, (\cD_\nu\, \chi_{J})   \\
 &~+~  \coeff{1}{16}\,e \, g^{\mu \nu} \,  \big( m^{IK} \, \cD_\mu\, m_{KJ}  \big)   \big( m^{JL} \, \cD_\nu\, m_{LI}  \big) ~-~   \coeff{1}{8}\, e \, g^{\mu \rho}  \, g^{\nu \sigma} \, m_{IK} \,m_{JL}\,  F_{\mu \nu }^{IJ}  \, F_{\rho \sigma }^{KL}   \\
& ~+~  \coeff{1}{2}\,e  \, \varepsilon^{\mu \nu \rho} \, \Big[  g_0 \,\big(A_\mu{}^{IJ}\, \partial_\nu  \widetilde A_\rho{}^{IJ}  ~+~\coeff{4}{3}\,  g_0 \, A_\mu{}^{IJ} \,  A_\nu{}^{JK}\, A_\rho{}^{KI} \,\big) ~+~  \coeff{1}{8}\,  {Y_\mu}{}^{IJ}  \, F_{\nu \rho}^{IJ} \Big]\\  
&~-~ 
\coeff{1}{2}\,e\, \cP_\mu^{Ir} \, \overline{\chi}^{\dot{A} r} \, \Gamma^I_{A\dot{A}}\gamma^\nu\gamma^\mu\psi^A_\nu
~+~ \coeff{1}{2}\,g\, e \, A_1^{AB} \,\overline{\psi}{}^A_\mu \gamma^{\mu\nu} \psi^B_\nu \\
&  
~+~  i\,g\,e \,A_2^{A\dot{A} r} \,\overline{\chi}^{\dot{A} r} \gamma^{\mu} \psi^A_\mu
~+~ \coeff{1}{2}\,g\,e \,A_3^{\dot{A} r\, \dot{B} s}\, \overline{\chi}^{\dot{A} r}\chi^{\dot{B} s}  ~-~e\, V  \,,
\end{aligned}
\label{eq:3Daction}
\end{equation}

The scalar potential is given by:
\begin{equation}
V ~=~  \coeff{1}{4}\, g_0^2   \,  \det\big(m^{IJ}\big) \, \Big [\, 2 \,\big(1- \coeff{1}{4} \,  (\chi_I \chi_I)\big)^2    ~+~ m_{IJ} m_{IJ}  ~+~\coeff{1}{2} \,  m_{IJ} \chi_I \chi_J  ~-~\coeff{1}{2} \,  m_{II}  \,  m_{JJ}\, \Big]
 \,,
\label{potential1}
\end{equation}
and our conventions for the $2\times 2$ space-time gamma matrices, $\gamma^\mu$, are given in Appendix \ref{sec:st-conventions}.

It is frequently convenient to  fix the local $SO(4)$ gauge symmetry by diagonalizing $P$ in terms of four scalar fields, $\muscal_i$:
\begin{equation}
P  ~=~  {\rm diag} \big(\,  e^{\muscal_1} \,, \,  e^{\muscal_2} \,, \,  e^{\muscal_3} \,, \,  e^{\muscal_4} \, \big) \,.
\label{Pdiag}
\end{equation}
The potential then reduces to:
\begin{equation}
\begin{aligned}
V~=~ & \coeff{1}{4}\, g_0^2   \,e^{-2\, (\muscal_1 +\muscal_2+\muscal_3+\muscal_4)} \, \Big [\, 2 \,\big(1- \coeff{1}{4} \,  (\chi_I \chi_I)\big)^2    ~+~ \big( e^{4\, \muscal_1}+e^{4\, \muscal_2}+e^{4\, \muscal_3}+e^{4\, \muscal_4}  \big)  \\
& \qquad\qquad\qquad\qquad \qquad\qquad~+~\coeff{1}{2}\, \big( e^{2\, \muscal_1}\, \chi_1^2 +e^{2\, \muscal_2}\, \chi_2^2 + e^{2\, \muscal_3}\, \chi_3^2+e^{2\, \muscal_4}\, \chi_4^2  \big) \\
& \qquad\qquad\qquad\qquad \qquad\qquad~-~\coeff{1}{2}\, \big( e^{2\, \muscal_1} +e^{2\, \muscal_2} + e^{2\, \muscal_3} + e^{2\, \muscal_4}  \big)^2  \, \Big] 
 \,.
\end{aligned}
\label{potential2}
\end{equation}

One can  use (\ref{cs_definition}), (\ref{ScalarDers}) and  (\ref{Pdiag}) to verify that the scalar and vector  kinetic terms are also given by:
\begin{equation}
\begin{aligned}
\cP_\mu^{Ir}  \,\cP^\mu{}^{\,Ir}  ~=~   g^{\mu \nu} \,\Big[ & \,  \coeff{1}{4} \, \big( m^{IK} \, \cD_\mu\, m_{KJ}  \big)   \big( m^{JL} \, \cD_\nu\, m_{LI}  \big) \\
  &  ~+~\coeff{1}{2} \, m^{IJ}   \, (\cD_\mu\, \chi_{I})  \, (\cD_\nu\, \chi_{J})   ~+~  \coeff{1}{4} \, \big( m^{IJ} \, m^{KL} \, C_{\mu}^{IK}  \, C_{\nu}^{JL}   \big) \, \Big] \,,
\end{aligned}
\label{scalkin}
\end{equation}
where $r = 1,\dots,5$.  See   \cite{Houppe:2020oqp} for more details.

The supersymmetry means that there is a superpotential
\begin{equation}
\begin{aligned}
W  ~\equiv~ & \coeff{1}{4} \, g_0 \,  (\det(P))^{-1}  \,   \Big [\, 2 \,\Big(1- \coeff{1}{4} \,  (\chi_A \chi_A)\Big) ~-~ {\rm Tr}\big(P\, P^T\big)  \, \Big] \\
~=~ & \coeff{1}{4} \, g_0 \,e^{-\muscal_1 -\muscal_2-\muscal_3-\muscal_4} \, \Big [\, 2 \,\Big(1- \coeff{1}{4} \,  (\chi_A \chi_A)\Big) ~-~ \Big( e^{2\, \muscal_1}+e^{2\, \muscal_2}+e^{2\, \muscal_3}+e^{2\, \muscal_4}  \Big) \, \Big]   \,,
 \end{aligned}
\label{superpot}
\end{equation}
and that the potential may be written as
\begin{equation}
V~=~ \delta^{ij} \frac{\partial W}{\partial \muscal_i}  \frac{\partial W}{\partial \muscal_j}  ~+~ 2\, m^{IJ} \, \frac{\partial W}{\partial \chi_I} \frac{\partial W}{\partial \chi_J}   ~-~2\, W^2  \,.
\label{potential3}
\end{equation}

The potential has a supersymmetric critical point\footnote{There are also non-supersymmetric flat directions extending from this supersymmetric critical point.} for  $\muscal_j = \chi_I =0$.   At this point $V$ takes the value
\begin{equation}
V_0 ~=~    -  \coeff{1}{2}\, g_0^2   \,.
\label{susypt}
\end{equation}
Setting all the other fields to zero, the Einstein equations give: 
\begin{equation}
R_{\mu \nu} ~=~ - 4 \, V_0 \, g_{\mu \nu} ~=~    2 \, g_0^2 \, g_{\mu \nu}  \,.
\label{susyvac}
\end{equation}
and the supersymmetric vacuum\footnote{One should note that because we are using a metric signature $(+ - - )$ the cosmological constant of AdS is positive, contrary to the more standard and rational choice of signature.} is an AdS$_3$ of radius, $g_0^{-1}$.  We therefore define
\begin{equation}
 \RAdS ~=~ \frac{1}{g_0} \,,
\label{RAdSscale}
\end{equation}
and so we will henceforth use this to set the overall scale of the metric.

%%%%%%%%%%%%%%%%%%%
\subsection{The supersymmetries and the BPS equations}
\label{sec:susies}
%%%%%%%%%%%%%%%%%%%

For supersymmetric backgrounds we must require $\delta\psi^A_\mu   = \delta\chi^{\dot{A} r} = 0$, and these  are the ``BPS equations'':
\begin{equation}
\delta\psi^A_\mu ~=~  D_\mu \epsilon^A ~-~ i  \,A_1^{AB}\gamma_\mu\epsilon^B ~=~ 0\,,    \qquad \delta\chi^{\dot{A} r} ~=~  \coeff{1}{2}\,i \,\Gamma^I_{A\dot{A}}\gamma^\mu\epsilon^A \, \cP_\mu^{Ir}  ~-~  A_2^{A\dot{A} r}\epsilon^A ~=~ 0\,.
\label{fermsusy}
\end{equation}
Note that the indices $r,s,\dots$ have the range  $1,\dots,5$.  Our gamma matrix conventions are given in Appendix \ref{sec:st-conventions}.

The covariant derivative on $\epsilon^A$ is defined by:
\begin{equation}
D_\mu \, \epsilon^A ~=~    \partial_\mu \epsilon^A    ~+~ \coeff{1}{4}\,\omega_\mu{}^{ab}\,\gamma_{ab}\, \epsilon^A + \coeff{1}{4}\,\cQ_\mu^{IJ}\Gamma^{IJ}_{AB}\, \epsilon^B \,.
\label{Depsilon}
\end{equation}

The quantities  $\cQ_\mu^{IJ}$ and $\cP_\mu^{Ir} $ are defined  in (\ref{ScalarDers}) and, for $P$ in the diagonal gauge (\ref{Pdiag}), the  $A$-tensors are given by:
\begin{equation}
\begin{aligned}
A_1^{AB}   ~=~ W\, (\Gamma^{1234}\big)_{AB} \,, \qquad A_2^{A\dot A \, r}   ~=~ & \frac{\partial W}{\partial \muscal_r} \, (\Gamma^{1234}\,\Gamma^{r}\big)_{A \dot A} \,, \quad 1 \le r \le 4  \,, \\
\qquad A_2^{A\dot A \, 5}   ~=~ &- \sqrt{2}\, \sum_{j=1}^4 \,e^{\muscal_j} \, \frac{\partial W}{\partial \chi_j} \, (\Gamma^{1234}\,\Gamma^{j}\big)_{A \dot A}  \,,
\end{aligned}
\label{A1A2tens2}
\end{equation}
where $W$ is the superpotential (\ref{superpot}) and there is no sum on $r$ in the expression for $A_2^{A\dot A \, r}$.   To obtain the expression for $A_2$ in a general gauge one can simply make a covariant $SO(4)$ rotation on (\ref{A1A2tens2}).

Superstrata and three-charge black holes in six dimensions  preserve precisely the supersymmetries that satisfy \cite{Houppe:2020oqp}:
\begin{equation}
    \Big(\oneone ~-~ \gamma^{12}\, \Gamma^{12}  \Big)\, \epsilon ~=~  0 \,, \qquad      \Big(\oneone ~+~ \gamma^{12}\, \Gamma^{34} \Big)\, \epsilon ~=~  0 \,.
    \label{basic-proj}
\end{equation}
Note that, together, these two projections imply:
\begin{equation}
    \Big(\oneone ~-~ \Gamma^{1234}  \Big)\, \epsilon ~=~  0 \,, 
    \label{int-proj}
\end{equation}
which was anticipated in (\ref{SO8proj2}). Moreover, any two of the  projectors from (\ref{basic-proj}) and (\ref{int-proj})  imply the third.

%%%%%%%%%%%%%%%%%%%
\subsection{The metric}
\label{sec:metric}
%%%%%%%%%%%%%%%%%%%

The original superstrata were parametrized in terms of the usual double null coordinates, $(u,v)$, which are related to the time and circle coordinates, $(t,y)$, via:
\begin{equation}
u ~\equiv~\frac{1}{\sqrt{2}} \, \big( t ~-~y  \big) \,, \qquad  v ~\equiv~\frac{1}{\sqrt{2}} \, \big(t ~+~y )\,, 
\label{uvtyreln}
\end{equation}
where $y$ is periodically identified as 
\begin{equation}
y ~\equiv~ y ~+~ 2 \pi \,R_y\,.
\label{yperiod}
\end{equation}
It is also convenient to compactify the radial coordinate and make the other coordinates scale-free:
\begin{equation}
\xi ~=~\frac{r}{\sqrt{r^2+ a^2}} \,,  \qquad  \tau~=~ \frac{t}{R_y}\,  \,,  \qquad  \psi~=~ \frac{\sqrt{2}\, v }{R_y}\,, 
\label{xidef}
\end{equation}
where $ 0 \le \xi < 1$ and $\psi$ inherits the periodicity $\psi \equiv \psi + 2 \pi$ from (\ref{yperiod}).

As noted in \cite{Houppe:2020oqp,Ganchev:2021pgs}, the most general three-dimensional metric can then be recast in the form:
\begin{equation}
ds_{3}^{2}  ~=~  \RAdSsq \, \bigg[ \,\Omega_1^{2} \, \bigg(d \tau +   \frac{k}{(1- \xi^{2})} \, d\psi \bigg)^2~-~\,\frac{\Omega_0^{2}}{(1-\xi^{2} )^{2}} \, \big( d \xi^2 ~+~ \xi^2 \, d \psi^2 \big) \, \bigg] \,,
\label{genmet1}
\end{equation}
for three arbitrary functions $\Omega_0$,  $\Omega_1$ and $k$ of the three coordinates, $(\tau ,\xi,\psi)$.

If one returns to the coordinates $(t,r,v)$, one obtains the more canonical superstratum metric:
\begin{equation}
ds_{3}^{2}  ~=~   \RAdSsq \, \bigg[ \, \frac{\Omega_1^{2}}{R_y^2} \, \bigg(dt  + \frac{\sqrt{2}}{a^2} \, (r^2  + a^2 ) \, k  \, dv  \bigg)^2~-~  \Omega_0^{2}\,\bigg(\frac{dr^2}{r^2 + a^2} ~+~\frac{2}{R_y^2 \, a^4 } \,r^2\,(r^2 + a^2) \, dv^2 \bigg)  \, \bigg]\,.
\label{genmet2}
\end{equation}
If one further sets:
\begin{equation}
\Omega_0 ~=~ \Omega_1 ~=~ 1\,, \qquad  k ~=~ \xi^2    ~=~  \frac{  r^2 }{ (r^2+ a^2) }  \,,
\label{AdSvals}
\end{equation}
then (\ref{genmet2}) becomes the metric of global AdS$_3$:
\begin{equation}
ds_{3}^{2}  ~=~ \RAdSsq \, \bigg[ \,   \bigg(1 + \frac{r^2}{a^2} \bigg)\,  \bigg(\frac{dt}{R_y}\bigg)^2~-~  \frac{dr^2}{r^2 + a^2}  ~-~ \frac{r^2}{a^2}\,  \bigg(\frac{dy}{R_y}\bigg)^2 \, \bigg] \,.
\label{AdSmet}
\end{equation}
%

%%%%%%%%%%%%%%%%%%%%%%%%%%%%%%%%%%%%%
\section{Truncations}
\label{sec:truncations}
%%%%%%%%%%%%%%%%%%%%%%%%%%%%%%%%%%%%%
%\vspace{0.8cm}

%%%%%%%%%%%%%%%%%%%
\subsection{The Q-ball truncation}
\label{sub:qball}
%%%%%%%%%%%%%%%%%%%

There are many ways to truncate the degrees of freedom of the three-dimensional theory while retaining some very interesting physical solutions. In particular one can make the ``simplest'' truncation of   \cite{Ganchev:2021pgs}.  Indeed, we will  impose an  invariance under three $U(1)$'s :
\begin{itemize}
\item[(i)]  The internal $U(1)$ that rotates the indices $(3,4)$.
\item[(ii)] A time translation by $\tau\to \tau -\alpha $, accompanied by internal $U(1)$ rotation by $\alpha \, \omega$ in  the $(1,2)$-direction, for an arbitrary parameter,   $\alpha$, and some frequency, $\omega$.
\item[(iii)]  A $\psi$-translation  by $\psi \to \psi - \alpha$, accompanied by internal $U(1)$ rotation by $ p\, \alpha$ in  the $(1,2)$-direction,  for an arbitrary parameter, $\alpha$, and for some mode number, $p$.
\end{itemize}
For the fields, $\chi_I$, the first symmetry implies that
\begin{equation}
 \chi_3 ~=~  \chi_4 ~=~ 0 \,,   \label{trunc1}
\end{equation}
while the second and third reduce $\chi_1$ and $\chi_2$  to a single mode:
\begin{equation}
 \chi_1 + i \chi_2 ~=~  \frac{a}{\sqrt{r^2 + a^2}}\,   \nu(\xi)\, e^{i (p \psi + \omega \tau)}  ~=~  \sqrt{1- \xi^2}\,   \nu(\xi)\, e^{i (p \psi + \omega \tau)}   \,. \label{singlemode1}
\end{equation}

These symmetries mean that   $P_{I}{ }^{\, J}$ can be reduced to the form
\begin{equation}
P_{I}{ }^{\, J} ~=~ \begin{pmatrix}
e^{ \newmuscal_1} \, S_{2 \times 2}  &0 _{2 \times 2}  \\
0 _{2 \times 2}    &  e^{ \newmuscal_2} \, \oneone_{2 \times 2}    
\end{pmatrix}\,,
 \label{mmatformq}
\end{equation}
where
\begin{equation}
S ~=~ {\cal O}^T  \, \begin{pmatrix}
e^{ \newmuscal_0}  &0 \\
0   &e^{-\newmuscal_0}   
\end{pmatrix}\,  {\cal O} \,, 
\qquad 
 {\cal O}  ~=~  \begin{pmatrix}
\cos  \sigma  & \sin  \sigma \\
- \sin  \sigma  &\cos  \sigma   
\end{pmatrix}\,,
  \label{SOmatform}
\end{equation}
for some scalar fields, $\newmuscal_0$, $\newmuscal_1$, $\newmuscal_2$ and where
\begin{equation}
 \sigma  ~=~ (p \, \psi ~+~\omega \, t) \,. \label{singlemode2}
\end{equation}
As we noted above, the gauge invariance can be used to diagonalize $P_{I}{ }^{\, J}$ , and here this reduces to  the freedom to use one of the  $U(1)$ gauge invariances to set $\omega = p =0$.  Fixing the gauge in this way means that one removes the phase dependences from all the fields and has: 
\begin{equation}
\chi_1   ~=~   \frac{a}{\sqrt{r^2 + a^2}}\,   \nu \,, \quad  \chi_2= \chi_3 = \chi_4    ~=~ 0   \,, \qquad P  ~=~  {\rm diag} \big(\,  e^{\newmuscal_1+ \newmuscal_0} \,, \,  e^{\newmuscal_1- \newmuscal_0} \,, \,  e^{\newmuscal_2} \,, \,  e^{\newmuscal_2} \, \big) \,.
\label{nophase}
\end{equation}
It turns out that, for the purposes of disentangling the various contributions to the BPS equations, this diagonal gauge is less practical than (\ref{singlemode1})--(\ref{SOmatform}), as will be elucidated on later.

The first symmetry reduces the gauge fields to $\tilde A^{12}$ and $\tilde A^{34}$, while remaining symmetries mean that these fields can only depend on $\xi$. The gauge fields can therefore be reduced to\footnote{One can remove components along $d\xi$ by a gauge transformation.}:  
\begin{equation}
\tilde A^{12} ~=~ \frac{1}{g_0} \,\big[\,  \Phi_1(\xi)  \, d\tau ~+~  \Psi_1(\xi)  \, d\psi \, \big]\,, \qquad  \tilde A^{34} ~=~ \frac{1}{g_0} \,\big[\,\Phi_2(\xi)  \, d\tau ~+~  \Psi_2(\xi)  \, d\psi    \, \big] \,,
  \label{gauge_ansatz}
\end{equation}
where we have  introduced explicit factors of $g_0^{-1}$ so as to cancel the $g_0$'s in the minimal coupling and thus render the fields and interactions scale independent. 

We also define the potential differences:
\begin{equation}
V_{\Phi_i} ~\equiv~ \Phi_i(1) - \Phi_i(0) \,, \qquad V_{\Psi_i} ~\equiv~ \Psi_i(1) - \Psi_i(0) \,,
  \label{potdiffs}
\end{equation}
 and note that, for the single mode solutions that we consider, these are gauge invariants.

Finally, the time-translation and $\psi$-translational invariance means that $\Omega_0$,  $\Omega_1$  and  $k$, can only depend on $\xi$.  

The Ansatz thus involves eleven arbitrary functions of  one variable, $\xi$:
\begin{equation}
{\cal F} ~\equiv~ \big\{\, \nu  \,, \ \  \mu_0   \,, \ \  \mu_1  \,, \ \ \mu_2  \,,  \ \  \Phi_1 \,, \ \  \Psi_1   \,, \ \  \Phi_2  \,, \ \ \Psi_2  \,, \ \  \Omega_0  \,, \ \  \Omega_1    \,, \ \  k \, \big\} \,.
  \label{functionlist}
\end{equation}
One can easily verify that this Ansatz is consistent with the equations of motion. 

%%%%%%%%%%% 
\subsubsection{The reduced action}
\label{sub:reduced-action}
%%%%%%%%%%% 

The easiest way to express the equations of motion for this truncated system is to give the action from which they can be derived.  This Lagrangian was given in \cite{Ganchev:2021pgs} and here we simply catalog this result. 

The Lagrangian can be decomposed into pieces:
\begin{equation}
    \cL ~=~ \cL_{\text{gravity}} + \cL_\chi + \cL_m + \cL_A + \cL_{CS} + \cL_Y - \sqrt{g}\, V \ ,
    \label{lagrangian_full}
\end{equation}
and the explicit expressions are: 
\begingroup
\allowdisplaybreaks
\begin{align}
\cL_{\text{gravity}} ~&\equiv~ -\frac14 \sqrt{g}\ R  \nonumber \\
~&=~  \frac{\Omega_1}{8\, g_0\, \xi}\, \bigg[ \,\frac{\Omega_1^2}{\Omega_0^2}\, \bigg(k' + \frac{2\,\xi}{1- \xi^2} \, k\bigg)^2  - 4\, \xi \,  \bigg(\partial_\xi\bigg( \xi\,  \frac{\Omega_0'}{\Omega_0}\bigg) + \frac{1}{\Omega_1} \,\partial_\xi\Big( \xi\,   \Omega_1' \Big)  \bigg)  -  \frac{16\, \xi^2}{(1- \xi^2)^2} \,    \bigg]
\\
\cL_\chi ~&\equiv~ \frac18 \sqrt{g}\, g^{\mu\nu}\, (\cD_\mu \chi_I)\, m^{IJ}\, (\cD_\nu \chi_J)  \nonumber \\
~&=~  \frac{e^{2(\mu_0 - \mu_1)}}{8\, g_0\, \xi \,(1- \xi^2) \,\Omega_1}\,  \,\bigg[\,  \Gamma \, \nu^2  ~-~   \xi^2\,(1- \xi^2)    \,\Omega_1^2 \, e^{-4\,\mu_0} \, \Big( \partial_\xi \Big (\sqrt{1- \xi^2}\, \nu \Big) \Big)^2    \, \bigg]
\\
\cL_m ~&\equiv~ \frac1{16} \sqrt{g} \, g^{\mu\nu} \, \Tr(m^{-1} (\cD_\mu m) m^{-1} (\cD_\nu m)) \nonumber   \\
~&=~  \frac{1}{2\, g_0\, \xi \,(1- \xi^2)^2 \,\Omega_1}\,  \,\bigg[\,  \Gamma \, \sinh^2 2\, \mu_0  ~-~   \xi^2\,(1- \xi^2)^2    \,\Omega_1^2 \,  \, \big( (\mu_0')^2 + (\mu_1')^2 +(\mu_2')^2 \big)   \, \bigg]
\\
\cL_A ~&\equiv~   -   \frac{1}{8}\, e \, g^{\mu \rho}  \, g^{\nu \sigma} \, m_{IK} \,m_{JL}\,  F_{\mu \nu }^{IJ}  \, F_{\rho \sigma }^{KL}  \nonumber  \\
~&=~  \frac1{2 g_0\, \xi\, \Omega_1} \Big[ \xi^2 \,\qty(e^{4 \mu_2}\ \Phi_1'^2 \,+\, e^{4\mu_1}\ \Phi_2'^2) \nonumber \\[-.3em]
& \qquad\qquad \qquad   -  \frac{\Omega_1^2}{\Omega_0^2}\,\qty(e^{4 \mu_2}\, \qty((1-\xi^2)\, \Psi_1' - k\, \Phi_1')^2 + e^{4\mu_1}\, \qty((1-\xi^2)\, \Psi_2' - k\, \Phi_2')^2) \Big]   
\\  
\cL_{CS} ~&\equiv~    \frac{1}{2}\,g_0 \, e  \, \varepsilon^{\mu \nu \rho} \,  \Big(A_\mu{}^{IJ}\, \partial_\nu  \widetilde A_\rho{}^{IJ}  ~+~\coeff{4}{3}\,  g_0 \, A_\mu{}^{IJ} \,  A_\nu{}^{JK}\, A_\rho{}^{KI} \,\Big)  \nonumber   \\
~&=~  \frac1{g_0} \qty( \Phi_1 \Psi_2' - \Psi_2 \Phi_1' + \Phi_2 \Psi_1' - \Psi_1 \Phi_2')  
\label{LCS}
\\
\cL_Y ~&\equiv~   \frac{1}{16}\,e  \, \varepsilon^{\mu \nu \rho} \,  {Y_\mu}{}^{IJ}  \, F_{\nu \rho}^{IJ}\nonumber \\
~&=~  \frac1{4g_0} \qty(1-\xi^2)\, \nu^2\,\qty((2 \Psi_1 + p)\, \Phi_2' \,-\, (2 \Phi_1 + \omega)\, \Psi_2')   \\
V ~&=~ \frac{g_0^2}{2}\, e^{-4(\mu_1 + \mu_2)} \Big[1 - 2\, e^{2(\mu_1 + \mu_2)} \cosh(2\mu_0) + e^{4\mu_1} \sinh^2(2\mu_0) \nonumber \\
&\qquad\qquad \qquad \qquad +   \frac1{16}  \nu^2\, \qty(1-\xi^2) \qty((1-\xi^2)\,\nu^2 + 4\, e^{2(\mu_0 + \mu_1)}  - 8) \Big] \,,
\end{align}%
\endgroup
where, $'$ indicates a differentiation with respect to $\xi$.  We are also using the convenient shorthand that captures the mode dependence and minimal couplings: 
\begin{equation}
 \Gamma ~=~ \xi^2  \Omega_0^2 \, \big[\,\omega + 2\, \Phi_1   \big]^2 ~-~ \Omega_1^2 \, \Big[\,(1-\xi^2 ) \,  \big(p + 2\,\Psi_1) ~-~k\, \big(\omega + 2\, \Phi_1    \big) \,\Big]^2 \,.
    \label{gauge_term}
\end{equation}
In particular, we note that the mode number, $p$,  and the frequency, $\omega$, can be absorbed into constant terms in $\Psi_1$ and $\Phi_1$, respectively.

There are also  three  integrals of the motion:
\begin{align}
H ~\equiv~ & \frac{\xi^2 \, (1- \xi^2)}{k} \, \bigg[\,  \frac{1}{\xi\,\Omega_1} \, \big(  \,  \widehat \cL_{\text{gravity}} + \cL_A  + \sqrt{g}\, V -   \cL_\chi -   \cL_m\,\big)  \nonumber\\
& \qquad \qquad\quad -\frac{1}{g_0 }\,  \,\bigg( (\mu_0')^2 + (\mu_1')^2 +(\mu_2')^2  ~+~\frac{1}{4}\,  e^{-2(\mu_0 + \mu_1)}\, \Big( \partial_\xi \Big (\sqrt{1- \xi^2}\, \nu \Big) \Big)^2  \bigg)   \, \bigg]  \,.
\label{Ham-constr}
\end{align}
and 
\begin{align}
\cI_1 ~\equiv~  & \frac{e^{4\,  \mu_1}\,(1- \xi^2)\,\Omega_1}{\xi\,\Omega_0^2} \, \big( (1- \xi^2) \, \Psi_2' - k\, \Phi_2' \big)~-~  \big(\omega + 2\, \Phi_1   \big)\, \big(1 - \coeff{1}{4}\, (1-\xi^2 ) \, \nu^2 \big) \,,  \label{Integral1}\\
\cI_2 ~\equiv~ &    \frac{e^{4\,  \mu_1}\, k\,\Omega_1}{\xi\,\Omega_0^2} \, \big( (1- \xi^2) \, \Psi_2' - k\, \Phi_2' \big) ~+~ \frac{e^{4\,  \mu_1}\,\xi \, \Phi_2' }{\Omega_1} ~-~  \big( p  + 2\, \Psi_1 \big)\, \big(1 - \coeff{1}{4}\, (1-\xi^2 ) \, \nu^2 \big) \,.\label{Integral2}
\end{align}
%

%%%%%%%%%%%%%%%%%%%
\subsection{The Coulomb branch} % (fold)
\label{sub:the_coulomb_branch}
%%%%%%%%%%%%%%%%%%%

The Coulomb branch may be thought of as the sector of the theory that is invariant under the $\ZZ_2$ symmetry: 
\begin{equation}
 \diag(-1, -1, -1, -1)  ~\in~ SO(4) 
 \end{equation}
 This invariance sets  $\chi_I = 0$ for all $I$, but retains the scalars in the matrix $P$ as well as all the gauge fields and symmetries.  The matrix, $P$, can thus be taken to be  diagonal.  In Section \ref{sec:the_coulomb_branch}, we will further simplify the problem by imposing a $U(1) \times U(1)$ symmetry which means that  $P$ has the form:
\begin{equation}
    P ~=~\diag (e^{\mu_1}, e^{\mu_1}, e^{\mu_2}, e^{\mu_2})
\end{equation}
The gauge fields and the metric are as before, (\ref{genmet1}) and (\ref{gauge_ansatz}), 
except,  that we will allow all the fields to depend on both $\xi$ and $\psi$.  Since we are seeking BPS solutions the fields cannot depend upon time, $t$.

%%%%%%%%%%% 
\subsubsection{The Lagrangian and the equations of motion} % (fold)
\label{ssub:the_lagrangian_and_the_equations_of_motion}
%%%%%%%%%%% 

The Lagrangian can once again be specialized to this field configuration, and we find:
\begin{equation}
\begin{aligned}
    \mathcal{L} ~&=~ - \frac{1}{4}  \sqrt{g} R \ +\ \frac{e}{2} g^{\mu\nu} \qty((\partial_\mu \mu_1)(\partial_\nu \mu_1) \,+\, (\partial_\mu \mu_2)(\partial_\nu \mu_2))  \ -\  \sqrt{g} \, V\\
    &\  - \frac{\sqrt{g}}{4} \qty(e^{4\mu_1} \tilde F_{\mu\nu}^{34}\,(\tilde F^{34})^{\mu\nu} \, +\, e^{4\mu_2} \tilde F_{\mu\nu}^{12}\,(\tilde F^{12})^{\mu\nu}) \ + \ \frac{g_0}{2} \epsilon^{\mu\nu\rho} \,\qty( \tilde A_\mu^{12} \,\tilde F_{\nu\rho}^{34} \,+\, \tilde A_\mu^{34} \,\tilde F_{\nu\rho}^{12})
\end{aligned}
\end{equation}
where $R$ is the Ricci scalar and $\epsilon$ is defined in (\ref{epsdefn}).  The scalar potential, (\ref{potential2}), reduces to:
\begin{equation}
    V ~=~ \frac{g_0^2}{2} \qty( e^{-4\, \left(\mu _1+\mu _2\right)}  - 2\, e^{-2\, \left(\mu _1+\mu _2\right)}) \,.
\end{equation}

On this part of the Coulomb branch, the BPS equations are generically insufficient to determine the solution and we will need (at least some of) the equations of motion. 
The scalar equations are:
\begin{equation}
\begin{aligned}
    \Delta \mu_1 ~&=~ - e^{4\mu_1} \,\tilde F_{\mu\nu}^{34}\,(\tilde F^{34})^{\mu\nu} \ +\ 2 \,g_0^2 \,\qty(e^{-4\mu_1 - 4 \mu_2} - e^{-2 \mu_1 - 2 \mu_2}) \\
    \Delta \mu_2 ~&=~ - e^{4\mu_2} \,\tilde F_{\mu\nu}^{12}\,(\tilde F^{12})^{\mu\nu} \ +\ 2 \,g_0^2 \,\qty(e^{-4\mu_1 - 4 \mu_2} - e^{-2 \mu_1 - 2 \mu_2}) \,.
\end{aligned}
\label{eq:eom_scalar}
\end{equation}
The Maxwell equations are:
\begin{equation}
    \begin{aligned}
        \partial_v \qty( \sqrt{g} \,e^{4\mu_2}\, \tilde F_{\mu\nu}^{12}) ~&=~ g_0\, \epsilon^{\mu\nu\rho}\, \tilde F_{\mu\nu}^{34} \\
        \partial_v \qty( \sqrt{g} \,e^{4\mu_1}\, \tilde F_{\mu\nu}^{34}) ~&=~ g_0\, \epsilon^{\mu\nu\rho}\, \tilde F_{\mu\nu}^{12} \,.
    \end{aligned}
    \label{eq:eom_maxwell}
\end{equation}
and Einstein equations read:
\begin{equation}
\begin{aligned}
    R_{\mu\nu} ~&=~ 2 \qty((\partial_\mu \mu_1)(\partial_\nu \mu_1) \,+\, (\partial_\mu \mu_2)(\partial_\nu \mu_2)) \,-\, 2 \qty(e^{4\mu_1} \tilde F_{\mu\rho}^{34}(\tilde F^{34})\indices{_\nu^\rho} \,+\, e^{4\mu_2} \tilde F_{\mu\rho}^{12}(\tilde F^{12})\indices{_\nu^\rho}) \\
    &\  + \qty(  2 \,g_0^2 \, \qty(2 e^{-2 \mu_1 - 2 \,\mu_2} - e^{-4 \mu_1 - 4 \mu_2}) \,+\, e^{4\mu_1} \tilde F_{\rho\sigma}^{34}\,(\tilde F^{34})^{\rho\sigma} \,+\, e^{4\mu_2} \tilde F_{\rho\sigma}^{12}\,(\tilde F^{12})^{\rho\sigma}) g_{\mu\nu} \,.
\end{aligned}
    \label{eq:eom_einstein}
\end{equation}
%

%%%%%%%%%%% 
\subsection{The superstratum solution} % (fold)
\label{ss:superstratum}
%%%%%%%%%%% 

It is  useful to note that the ``single-mode superstratum''  corresponds to \cite{Mayerson:2020tcl, Ganchev:2021pgs}: 
\begin{equation}
\begin{aligned}
\nu   ~=~&\alpha_0 \,   \xi^n  \,, \qquad  \mu_1 ~=~ \coeff{1}{2}\,  \log \Big[ \, 1 - \coeff{1}{4}\, \alpha_0^2 \,  (1-\xi^2)\, \xi^{2n}  \, \Big]   \,,\qquad     \mu_0 ~=~ \mu_2 ~=~ 0 \,,  \\
  \Phi_1  ~=~  &\frac{1}{2}  \,, \qquad \Psi_1  ~=~ 0  \,,  \\  
  \Phi_2 ~=~ &\frac{1}{2} \,\bigg[\,1 ~-~  \frac{1}{\big(\,1 - \coeff{1}{4}\, \alpha_0^2 \, (1-\xi^2)\,  \xi^{2n} \, \big)}  \, \bigg]\,, \qquad  \Psi_2 ~=~
-   \frac{\alpha_0^2 }{8}  \,\frac{ \xi^{2n+2}}{\big(\, 1 - \coeff{1}{4}\, \alpha_0^2 \,  (1-\xi^2)\, \xi^{2n} \, \big)} \,,\\ 
  \Omega_0  ~=~ &  \sqrt{\, 1 - \coeff{1}{4}\, \alpha_0^2 \,  (1-\xi^2)\, \xi^{2n} }     \,, \qquad  \Omega_1 ~=~  1    \,, \qquad   k ~=~\xi^2  \,,
\end{aligned}
  \label{ssres1}
\end{equation}
where we have made a gauge choices for $ \tilde A^{34}$, (\ref{gauge_ansatz}), so that $\Phi_2(1) =0$ and $\Psi_2(0) = 0$.

%%%%%%%%%%%%%%%%%%%%%%%%%%%%%%%%%%%%%
\section{Preparing the computation }
\label{sec:preparing_computation}
%%%%%%%%%%%%%%%%%%%%%%%%%%%%%%%%%%%%%
%\vspace{0.8cm}

We will set up the BPS equations in a manner that they will cover both the Q-ball truncation and the Coulomb branch.  In particular, this means that we will initially allow many of the fields to depend on $\xi$ and $\psi$.  We will require the solutions to be time-independent and so we choose the gauge $\omega =0$ for the Q-ball truncation.  

We will also simplify the BPS computation by taking   $\Omega_1 \equiv 1$. This choice is consistent with the original superstratum \cite{Mayerson:2020tcl,Houppe:2020oqp}  and is predicted for the generalized superstratum  \cite{Ganchev:2021pgs}.   However, this choice does involve modifying the relative normalization of $\tau$ given in (\ref{xidef}) by: 
\begin{equation}
\tau ~=~ a^2 g_0^4 \, R_y t \,.
\end{equation}
It may be possible to find Coulomb branch solutions with non-constant $\Omega_1$, but we will not investigate this possibility here as we can find non-trivial results for $\Omega_1 \equiv 1$.   Similarly, we will also assume $\partial_\psi k = 0$: again this is true of the superstratum and predicted tor the generalized superstratum.  We will also make this assumption for the Coulomb branch solutions. 

The field,  $\nu$, vanishes in the Coulomb branch flows and only depends on $\xi$ in the Q-ball truncation, and so we will also restrict   $\nu $ to being a function of $\xi$ alone throughout this analysis.   The remaining fields, $\Omega_0$, $\mu_0$,  $\mu_1$,  $\mu_2$,  $\Phi_j$,   $\Psi_j$ will, initially, be allowed to be functions of both $\xi$ and $\psi$.  We will specialize them later to the two classes of truncation. 

With these assumptions, we now catalog all the terms that go into the supersymmetry variations and thus define the BPS equations.

%%%%%%%%%%%%%%%%%%%
\subsection{The frames and connection}
\label{sub:frames}
%%%%%%%%%%%%%%%%%%%

We use the frames that are best adapted to the supersymmetry:
\begin{equation}
    e^0 ~=~ g_0^{-1} \qty(d \tau +   \frac{k}{(1- \xi^{2})} \, d\psi) \,, \qquad 
    e^1 ~=~ g_0^{-1} \frac{\Omega_0}{1-\xi^{2}} \, d \xi \,, \qquad  
    e^2 ~ =~ g_0^{-1} \frac{\Omega_0}{1-\xi^{2}} \, \xi \, d\psi
    \label{eq:frames}
\end{equation} 
The spin connections,  defined as $de^a + \omega^a_b \wedge e^b = 0$,   are given by:
\begin{equation}
    \omega_{01} ~=~ s_0 e^2 \ , \quad \omega_{02} ~=~ -s_0 e^1 \qand \omega_{12} ~=~ -s_0 e^0 + s_1 e^1 + s_2 e^2
    \label{eq:spin_connections}
\end{equation}
where
\begin{equation}
\begin{aligned}
    s_0 ~=~ &  g_0 \frac{2\xi k + (1-\xi^2) \partial_\xi k}{2\xi \Omega_0^2} \,, \qquad \qquad 
    s_1 ~=~  -g_0\frac{(1-\xi^2) \partial_\psi\Omega_0}{\xi\Omega_0^2}  \,  \\ 
    s_2 ~=~& g_0\frac{(1+\xi^2) \Omega_0 + \xi (1-\xi^2) \partial_\xi\Omega_0}{\xi\Omega_0^2} \,.
\end{aligned}
    \label{eq:connections_values}
\end{equation}

%%%%%%%%%%%%%%%%%%%
\subsection{The scalar terms} % (fold)
\label{sub:scalar_terms}
%%%%%%%%%%%%%%%%%%%

The scalar kinetic terms, $\cP$,  defined in (\ref{ScalarDers}),  are given by:
\begin{align}
    \begin{split}
        \cP^{IJ} &= g_0 \sinh(2\mu_0) \qty(2 \Phi_1 e^0 + \frac{(1-\xi^2)(p + 2\Psi_1) - 2 k \Phi_1}{\xi\Omega_0} e^2) \cO_1^{IJ} \\
        &\qquad + (d\mu_0) \cO_2^{IJ} + \diag(d\mu_1, d\mu_1, d\mu_2, d\mu_2) - \frac12 \mathscr{C}^{IJ}
    \end{split}\label{eq:scalP}
    \\
    \begin{split}
        g_0^{-1} \cP^{I5} &= -e^{\mu_0 - \mu_1} \sqrt{\frac{1-\xi^2}2} \nu \qty(2 \Phi_1 e^0 + \frac{(1-\xi^2)(p + 2\Psi_1) - 2 k \Phi_1}{\xi\Omega_0} e^2) \cO_1^{I1} \\
        &\qquad + e^{-\mu_0 - \mu_1} \frac1{\sqrt{2}} d\qty(\sqrt{1-\xi^2} \nu) \cO_2^{I1}
    \end{split} \label{eq:scalP5}
\end{align}
where
\begin{equation}
    \cO_1 ~=~ \pmqty{
    -\sin(2p\psi) & \cos(2p\psi) & 0 & 0 \\
    \cos(2p\psi) & \sin(2p\psi) & 0 & 0 \\
    0 & 0 & 0 & 0 \\
    0 & 0 & 0 & 0}
    \qand 
    \cO_2 ~=~ \pmqty{
    \cos(2p\psi) & \sin(2p\psi) & 0 & 0 \\
    \sin(2p\psi) & -\cos(2p\psi) & 0 & 0 \\
    0 & 0 & 0 & 0 \\
    0 & 0 & 0 & 0}
\end{equation}

The composite connection, $\cQ$, is also  defined in (\ref{ScalarDers}) and is given by:
\begin{align}
    \cQ^{12} ~&=~ -g_0\cosh(2\mu_0) \qty(2 \Phi_1 e^0 + \frac{(1-\xi^2)(p + 2\Psi_1) - 2 k \Phi_1}{\xi\Omega_0} e^2) \\
    &\qquad + g_0\frac{p(1-\xi^2)}{\xi\Omega_0} e^2 + \frac12 \mathscr{C}^{12} 
    \\
    \cQ^{34} ~&=~ - g_0 \qty(2 \Phi_2 e^0 + \frac{2(1-\xi^2)\Psi_2 - 2 k \Phi_2}{\xi\Omega_0} e^2) + \frac12 \mathscr{C}^{34}
    \label{eq:scalQ}
\end{align}

The non-zero components of the Chern-Simons term (\ref{cs_definition}) are given by:
\begin{align}
    \mathscr{C}^{12} ~&=~ \frac{2 g_0 e^{2\mu_1} (1 - \xi^2)}{\Omega_0} \qty(\frac{(1-\xi^2) \partial_\xi\Psi_2 - k \partial_\xi\Phi_2}{\xi \Omega_0} e^0 \ -\ \frac1\xi (\partial_\psi \Phi_2) e^1 \ +\ (\partial_\xi \Phi_2) e^2) \\
    \mathscr{C}^{34} ~&=~ \frac{2 g_0 e^{2\mu_2} (1 - \xi^2)}{\Omega_0} \qty(\frac{(1-\xi^2) \partial_\xi\Psi_1 - k \partial_\xi\Phi_1}{\xi \Omega_0} e^0 \ -\ \frac1\xi (\partial_\psi \Phi_1) e^1 \ +\ (\partial_\xi \Phi_1) e^2)
    \label{eq:chern_simons}
\end{align}
%

%%%%%%%%%%%%%%%%%%%
\subsection{The A-tensors} % (fold)
\label{sub:the_a_tensors}
%%%%%%%%%%%%%%%%%%%

The $A$-tensors were defined in (\ref{A1A2tens2}) for a diagonal matrix of scalars, $P$. While one can always choose such a gauge, it is somewhat more convenient to work in the gauge defined by  (\ref{singlemode1}), (\ref{mmatformq}) and  (\ref{SOmatform}). Obviously a choice of gauge cannot modify the complete set of BPS equations, however, working in the more general gauge simplifies the extraction of all the independent pieces of the BPS system. To work in this more general gauge we need to make the appropriate $U(1)$ gauge transformations of the $A$-tensors  in (\ref{A1A2tens2}).

The expression for  the superpotential in the first line of   (\ref{superpot}) is manifestly gauge invariant.  For the scalars we are considering, the superpotential  becomes:
\begin{equation}
    W ~=~ \frac12 g_0 \, e^{-2 \mu_1 - 2 \mu_2} \qty(1 \,-\, \frac14 (1-\xi^2) \nu^2 \,-\, e^{2\mu_1} \cosh(2\mu_0) \,-\, e^{2\mu_2}) \,.
    \label{eq:superpotential}
\end{equation}

The tensors are then given by:
\begin{equation}
    A_1^{AB} ~=~ W \qty(\Gamma^{1234})_{AB} \ ,\quad
    A_2^{A \dot{A} 5} ~=~ - \sqrt{2} P\indices{_I^J} \pdv{W}{\chi_J} \, \qty(\Gamma^{1234} \Gamma^I)_{A\dot{A}}
    \label{eq:atensors_1}
\end{equation}
and
\begin{equation}
    A_2^{A \dot{A}r} ~=~ \left\{\ 
    \begin{aligned}
        &\frac12 \qty[(\partial_{\mu_1}W) \delta^{Ir} + (\partial_{\mu_0}W) \cO_2^{Ir}] \qty(\Gamma^{1234} \Gamma^I)_{A\dot{A}} & \qq*{if} 1 \le &r \le 2
        \\
        &\frac12 (\partial_{\mu_2}W) \qty(\Gamma^{1234} \Gamma^r)_{A\dot{A}} & \qq*{if} 3 \le &r \le 4
    \end{aligned} \right.
    \label{eq:atensors_2}
\end{equation}

Note that the factors $\frac{1}{2}$, compared to  the previous definition, (\ref{A1A2tens2}), arise because of the double-degeneracy of the eigenvalues of $P$.

%%%%%%%%%%%%%%%%%%%%%%%%%%%%%%%%%%%%%
\section{The BPS equations} % (fold)
\label{sec:the_susy_equations}
 %%%%%%%%%%%%%%%%%%%%%%%%%%%%%%%%%%%%%

Our purpose here is not an exhaustive classification of BPS solutions, but to get to the generalized superstrata, and so we will occasionally take a  short cut that could potentially omit  some solutions but will get us efficiently to our target.

From here on, when the indices on the various matrices are omitted, it is to be understood that there is implicit  left-multiplication.   

%%%%%%%%%%%%%%%%%%%
\subsection{The first set of equations} % (fold)
\label{sub:the_first_equation}
%%%%%%%%%%%%%%%%%%%

Using (\ref{eq:atensors_1}), the equation for the fifth component of the fermion variation (\ref{fermsusy}) is:
\begin{equation}
\frac12 i \,\Gamma^I_{A\dot{A}}\gamma^\mu\epsilon^A \, \cP_\mu^{I5}  ~- ~ \sqrt{2} (P \, \partial_\chi W)_I \, (\Gamma^{1234}\,\Gamma^{I}\big)_{A \dot A} \epsilon^A ~=~ 0\,,
\label{eq:firstBPS}
\end{equation}
The sum over $I$ runs from $1$ to $2$, since the elements $3$ and $4$ are trivial.

We can multiply (\ref{eq:firstBPS}) by $\gamma^1$ to the left, and $\Gamma^2_{\dot{A}B}$ to obtain:
\begin{equation}
\begin{aligned}
    &\frac12 i \qty(\Gamma^{12} \cP_1^{15} - \Gamma^{12} \gamma^{12} \cP_2^{15} - \cP_1^{25} + \gamma^{12} \cP_2^{25}) \epsilon \\
    &+ \frac12 \gamma^{12} \qty(\Gamma^{12} \cP_0^{15} - \cP_0^{25}) \epsilon + \qty( \sqrt{2} (P \partial_\chi W)_1 \Gamma^{12} \Gamma^{1234} - \sqrt{2} (P \partial_\chi W)_2 \Gamma^{1234} )\epsilon ~=~  0 \,.
\end{aligned}
\end{equation}
One can rewrite this as:
\begin{equation}
\begin{aligned}
    &\frac12 i \qty(\Gamma^{12} \cP_1^{15} + \gamma^{12} \cP_2^{25}) \epsilon - \frac12 i \qty(\cP_1^{25} + \Gamma^{12} \gamma^{12} \cP_2^{15})\epsilon \\
    &+ \qty(\frac12 \gamma^{12} \Gamma^{12} \cP_0^{15} - \sqrt{2} (P \partial_\chi W)_2 \Gamma^{1234} ) \epsilon - \qty(\frac12 \gamma^{12} \cP_0^{25} - \sqrt{2} (P \partial_\chi W)_1 \Gamma^{12}\Gamma^{1234} ) \epsilon  ~=~  0 \,.
\end{aligned}
\label{eqn:BPS1} 
\end{equation}

After applying the projection conditions (\ref{basic-proj}), one sees that this equation is solved\footnote{As noted above, this is one of our ``short cuts:'' there may be more general ways to satisfy (\ref{eqn:BPS1}), but we will not explore this possibility here.} if one requires:
\begin{align}
    \cP_0^{15} ~&=~ 2\sqrt{2} (P \partial_\chi W)_2 \,, \qquad
    \cP_0^{25} ~=~ - 2\sqrt{2} (P \partial_\chi W)_1 \,, \\
    \cP_1^{15} ~&=~ \cP_2^{25}  \,, \qquad \qquad \qquad
    \cP_2^{15} ~=~ - \cP_1^{25} \,,
\end{align}
where the lower indices are frame indices.

For the configurations we consider, the first two conditions are identical, as  are the last two; and this is actually the only way to solve theses equations given  the projection conditions. Using (\ref{eq:scalP5}) and (\ref{eq:superpotential}), the first two conditions lead to the equation:
\begin{equation}
    \nu \, \qty(\Phi_1 - \frac12 e^{-2\mu_2}) ~=~ 0 \,,
    \label{eq:phi_1}
\end{equation}
while the last two lead to:
\begin{equation}
    \xi (1-\xi^2) \partial_\xi \nu ~=~ \qty[\xi^2 + e^{2 \mu_0} \qty((1-\xi^2)(p + 2 \Psi_1) - 2 k \Phi_1)]\nu \,. \label{eq:nu_first}
\end{equation}
%

%%%%%%%%%%%%%%%%%%%
\subsection{The second set of equations} % (fold)
\label{sub:the_second_set_of_equations}
%%%%%%%%%%%%%%%%%%%

The equations for the other fermion variations (\ref{fermsusy}) are:
\begin{equation}
\frac12 i \,\Gamma^I_{A\dot{A}}\gamma^\mu\epsilon^A \, \cP_\mu^{Ir}  ~- ~ A_2^{A \dot A r} \epsilon^A ~=~ 0\,,
\label{eq:secondBPS}
\end{equation}
where $r$ runs from $1$ to $4$.

We will see that the equations for $r=1,2$ are equivalent, as are the equations for $r=3,4$.

%%%%%%%%%%% 
\subsubsection{Equations from $r=3,4$} % (fold)
\label{ss:for_r3}
%%%%%%%%%%% 

We start with $r=3$, and use (\ref{eq:atensors_2}).  After multiplying by $\Gamma^4$ on the right, and $\gamma^2$ on the left, and rearranging the terms, we find:
\begin{equation}
    \frac12 i \qty(\Gamma^{34} \cP_1^{33} - \gamma^{12} \cP_2^{34}) \epsilon + \frac12 i \qty(\cP_1^{34} - \Gamma^{34} \gamma^{34} \cP_2^{33})\epsilon + \frac12 \qty(\gamma^{12} \cP_0^{34} + \qty(\partial_{\mu_2} W) \Gamma^{34}\Gamma^{1234} ) \epsilon ~=~ 0 \,.
 \label{r3susy}
\end{equation}

An obvious way to solve this equation is to  require that the three terms in parentheses vanish independently\footnote{As noted above, we are making ``short cuts:'' there may be more general ways to satisfy (\ref{r3susy}) and  (\ref{eqn:BPS2}), but we will not explore this possibility here \label{footnoteshortcut}.}. Applying  the projection conditions (\ref{basic-proj}) these three equations lead to:
\begin{equation}
    \cP_1^{33} \ =\ \cP_2^{34} \ ,\quad \cP_1^{34} \ =\ -\cP_2^{33} \qand \cP_0^{34} \ =\ - \partial_{\mu_2} W \,.
\end{equation}

Using (\ref{eq:scalP}), the first equation  leads to
\begin{equation}
    \partial_\xi\mu_2 + e^{2\mu_2} \partial_\xi\Phi_1 ~=~ 0
    \label{eq:redundant1a}
\end{equation}
while the second leads to
\begin{equation}
    \partial_\psi\mu_2 + e^{2\mu_2} \partial_\psi\Phi_1 ~=~ 0
    \label{eq:redundant1b}
\end{equation}
If $\nu \ne 0$ then these  two equations are redundant by virtue of  (\ref{eq:phi_1}).  Thus, at least for $\nu\ne 0$, the only non-trivial term in (\ref{r3susy}) is the last parenthesis, and, using (\ref{eq:superpotential}), this leads to:
.%
\begin{equation}
    \frac{1-\xi^2}{\xi} \frac{e^{2\mu_1 + 4 \mu_2}}{\Omega_0^2} \qty(k \partial_\xi \Phi_1 - (1-\xi^2) \partial_\xi\Psi_1) ~=~ 1 - e^{2\mu_1} \cosh(2 \mu_0) - \frac{1-\xi^2}4 \nu^2
    \label{eq:phi_1_psi_1}
\end{equation}

As we noted earlier, the equation for with $r=4$  leads to the same set of equations.

%%%%%%%%%%% 
\subsubsection{Equations from $r=1,2$} % (fold)
\label{ss:for_r1}
%%%%%%%%%%% 

For $r=1$, we use (\ref{eq:atensors_2}) and  multiply to the right by $\Gamma^2$ and on the left by $\gamma^2$.  Rearranging the result then leads to:
\begin{equation}
\begin{aligned}
    &\frac12 i \qty(\Gamma^{12} \cP_1^{11} - \gamma^{12} \cP_2^{12}) \epsilon + \frac12 i \qty(\cP_1^{12} - \Gamma^{12} \gamma^{12} \cP_2^{11})\epsilon
    + \frac12 \qty(\gamma^{12} \Gamma^{12} \cP_0^{11} - \qty(\partial_{\mu_0} W) \sin(2p \psi) \Gamma^{1234} ) \epsilon
    \\
    &+ \frac12 \qty(\gamma^{12} \cP_0^{12} + \qty(\qty(\partial_{\mu_1} W) + \qty(\partial_{\mu_0} W) \cos(2p\psi)) \Gamma^{12}\Gamma^{1234} ) \epsilon = 0
\end{aligned}
\label{eqn:BPS2} 
\end{equation}

Once again we ask for each term to vanish independently$^{\ref{footnoteshortcut}}$. We apply the projection conditions (\ref{basic-proj}) and arrive at:
\begin{align}
     \cP_1^{11} ~&=~ -\cP_2^{12} \,, \qquad \qquad \qquad
     \cP_1^{12} ~ =~ \cP_2^{11} \\
     \cP_0^{11} ~&=~ \qty(\partial_{\mu_0} W) \sin(2p\psi)  \,, \qquad
     \cP_0^{12} ~=~ - \qty(\partial_{\mu_0} W) - \qty(\partial_{\mu_1} W) \cos(2p\psi)
 \end{align} 

Using (\ref{eq:scalP}) and (\ref{eq:superpotential}), we see that all these equations are of the form $A + B \cos(2p\psi) + C \sin(2p\psi) = 0$ for some $A,B,C$. We are going to assume$^{\ref{footnoteshortcut}}$  that this means $A = B = C = 0$.

The first and second equations lead to the simple constraints
\begin{equation}
    \partial_\xi \mu_1 - e^{2\mu_1} \partial_\xi \Phi_2 ~=~ 0 \ , \qquad \partial_\psi \mu_1 - e^{2\mu_1} \partial_\psi \Phi_2  ~=~ 0  \ , \qquad \partial_\psi \mu_0 ~=~ 0
    \label{eq:phi_2}
\end{equation}
along with the more interesting equation
\begin{equation}
    (1-\xi^2) \xi \partial_\xi \mu_0 ~=~ -\sinh(2 \mu_0) \qty[ 2 k \Phi_1 - (1-\xi^2) (p + 2 \Psi_1)]
    \label{eq:mu_0} 
\end{equation}

The third and fourth equations lead to the simple constraint 
\begin{equation}
    \sinh(\mu_0) \, \qty(\Phi_1 - \frac12 e^{-2\mu_2}) ~=~ 0
    \label{eq:redundant2}
\end{equation}
which is redundant with (\ref{eq:phi_1}) if $\nu\neq 0$ and $\mu_0 \neq 0$ ; along with the interesting equation :
\begin{equation}
    - \frac{1-\xi^2}{\xi} \frac{e^{4\mu_1 + 2 \mu_2}}{\Omega_0^2} \qty(k \partial_\xi \Phi_2 - (1-\xi^2) \partial_\xi\Psi_2) ~=~ 1 - e^{2\mu_2} - \frac{1-\xi^2}4 \nu^2 \,.
    \label{eq:phi_2_psi_2}
\end{equation}

Again, the equation for with $r=2$  leads to the same result.

%%%%%%%%%%%%%%%%%%%
\subsection{The third set of equations} % (fold)
\label{sub:the_third_set_of_equations}
%%%%%%%%%%%%%%%%%%%

The equation for the gravitino variation (\ref{fermsusy}) is
\begin{equation}
\dd \epsilon^A    ~+~ \frac{1}{4}\,\omega_{ab}\,\gamma^{ab}\, \epsilon^A + \frac{1}{4}\,\cQ^{IJ}\Gamma^{IJ}_{AB}\, \epsilon^B  ~-~ i W\, (\Gamma^{1234}\big)_{AB} \,\gamma_c \, e^c \,\epsilon^B ~=~ 0 \,.
\label{eq:thirdBPS}
\end{equation}

Using (\ref{eq:spin_connections}), we obtain
\begin{equation}
    \begin{aligned}
        \dd{\epsilon} \ &+\   \frac12 s_0 \qty(\gamma^{01} e^2 - \gamma^{02} e^1 - \gamma^{12} e^0)\, \epsilon \ +\  \frac12 (s_1 e^1 + s_2 e^2) \gamma^{12}\,\epsilon
        \\
        & + \frac{1}{4}\,\cQ^{IJ}\Gamma^{IJ}\, \epsilon \ - \  W (i \gamma_c e^c) \Gamma^{1234}\, \epsilon ~=~ 0 \,,
    \end{aligned}
\end{equation}
which we can rewrite  as:
\begin{equation}
\begin{aligned}
    \dd{\epsilon}  ~&=~ \qty(\frac12 s_0 + W \Gamma^{1234}) \qty(\gamma^{02} e^1 - \gamma^{01} e^2) \epsilon
    \\
    & - e^0 \qty(\frac12 \cQ_0^{12} \Gamma^{12} + \frac12 \cQ_0^{34} \Gamma^{34} - \frac12 s_0 \gamma^{12} + W \gamma^{12} \Gamma^{1234}) \epsilon
    \\
    & - e^1 \qty(\frac12 \cQ_1^{12} \Gamma^{12} + \frac12 s_1 \gamma^{12}) \epsilon
    \\
    & - e^2\, \qty(\frac12 \cQ_2^{12} \,\Gamma^{12} + \frac12 \cQ_2^{34} \,\Gamma^{34} + \frac12 s_2 \gamma^{12}) \epsilon \,,
\end{aligned}
\label{eq:bps_3_rewritten}
\end{equation}
where $s_0$ is defined in   (\ref{eq:connections_values}).

We allow $\epsilon$ to have a phase dependence on $t$ and $\psi$, and our $U(1)$ invariance means that this can be  induced via gauge rotation in the $(1,2)$ directions.  We therefore allow:
\begin{equation}
    \dd{\epsilon}  ~=~ (\eta_t \dd{t} + \eta_\psi \dd{\psi}) \Gamma^{12} \epsilon ~=~ g_0\qty(\eta_t e^0 + \frac{1}{\xi \Omega_0}\qty((1-\xi^2) \eta_\psi - k \eta_t) e^2) \Gamma^{12} \epsilon 
\end{equation}
for some gauge-dependent, real parameters, $\eta_t$ and $\eta_\psi$.  Note that, because of the projection condition  (\ref{int-proj}), the gauge rotation  $(1,2)$  direction also implicitly rotates $\epsilon$ in the $(3,4)$ directions, and so $\eta_t$ and $\eta_\psi$ will depend on all the $U(1)$ gauge choices.

We now distribute the terms $\eta_t$ and $\eta_\psi$ to their logical place in equation (\ref{eq:bps_3_rewritten}), and ask that each term vanishes independently. After applying the projection conditions (\ref{basic-proj}), we get
\begin{align}
    W ~&=~ - \frac12 s_0 \\
    \frac12 \qty(\cQ_0^{12} - \cQ_0^{34}) ~&=~ -g_0 \eta_t + W - \frac12 s_0 ~=~ -g_0 \eta_t + 2 W\\
    \frac12 \qty(\cQ_1^{12} - \cQ_1^{34}) ~&=~ \frac12 s_1\\
    \frac12 \qty(\cQ_2^{12} - \cQ_2^{34}) ~&=~ - \frac{g_0}{\xi \Omega_0}\qty((1-\xi^2) \eta_\psi - k \eta_t)+ \frac12 s_2
\end{align}

We now use (\ref{eq:scalQ}), (\ref{eq:connections_values}) and (\ref{eq:superpotential}). The first equation leads to
\begin{equation}
    e^{2\mu_2} - \frac{e^{2\mu_1 + 2 \mu_2}}{2\xi \Omega_0^2} \qty((1-\xi^2) \partial_\xi k + 2\xi k) ~=~ 1 - e^{2\mu_1} \cosh(2 \mu_0) - \frac{1-\xi^2}4 \nu^2 \,.
    \label{eq:k}
\end{equation}

The second equation leads to
\begin{equation}
\begin{aligned}
    &\frac{1-\xi^2}{2\xi \Omega_0^2} \big[e^{2\mu_1} (k \partial_\xi\Phi_2 - (1-\xi^2) \partial_\xi\Psi_2) - e^{2\mu_2} (k \partial_\xi\Phi_1 - (1-\xi^2) \partial_\xi\Psi_1))\big] ~=~ \\
    &\eta_t + \Phi_2 - \cosh(2\mu_0) \Phi_1 + e^{-2(\mu_1+\mu_2)} \qty(-1 + e^{2\mu_1} \cosh(2\mu_0) + e^{2\mu_2} + \frac14 (1-\xi^2) \nu^2) \,.
\end{aligned}
\label{eq:redundant3}
\end{equation}

The third equation leads to
\begin{equation}
    \frac{\partial_\psi \Omega_0}{\Omega_0} \ +\ e^{2 \mu _2}  \partial_\psi\Phi _1 - e^{2 \mu _1}  \partial_\psi\Phi _2 ~=~ 0 \,.
    \label{eq:omega_0_psi}
\end{equation}

And the fourth equation leads to
\begin{equation}
    \begin{split}
    \xi \qty( \frac{\partial_\xi \Omega_0}{\Omega_0} + e^{2 \mu _2}  \partial_\xi\Phi _1-e^{2 \mu _1}  \partial_\xi\Phi _2) ~&=~ \frac{\cosh(2\mu_0)}{1-\xi^2} \qty(2 k \Phi _1 - (1-\xi^2) (p + 2 \Psi _1)) + p + 1\\
    &  - \frac{2}{1-\xi^2} \qty(k \qty(\eta_t + \Phi _2 ) - (1-\xi^2) (\eta_\psi + \Psi _2) + 1) \,.
    \end{split}
    \label{eq:omega_0}
\end{equation}

% Replacing everything leads to :
% \begin{equation}
%     \begin{aligned}
%         g_0^{-1} \dd{\epsilon} \ &+\  \frac{2\xi k + (1-\xi^2) k'}{4\xi \Omega_0^2}\qty(\gamma^{01} e^2 - \gamma^{02} e^1 - \gamma^{12} e^0)\, \epsilon \ +\  \frac{(1+\xi^2) \Omega_0 + \xi (1-\xi^2) \Omega_0'}{2\xi\Omega_0^2} e^2 \gamma^{12}\,\epsilon
%         \\
%         &-\ e^0 \qty(\cosh(2\mu_0) \Phi_1 \Gamma^{12} + \Phi_2 \Gamma^{34})\, \epsilon + \frac{p(1-\xi^2)}{2\xi \Omega_0}e^2 \Gamma^{12} \epsilon \ +\  \frac14 e^a \qty(\mathscr{C}_a^{12} \Gamma^{12} + \mathscr{C}_a^{34} \Gamma^{34})\, \epsilon
%         \\
%         &-\  \frac{1}{2\xi \Omega_0} e^2 \qty[\cosh(2\mu_0)\qty((1-\xi^2)(p + 2\Psi_1) - 2 k \Phi_1) \Gamma^{12} + \qty(2(1-\xi^2)\Psi_2 - 2 k \Phi_2) \Gamma^{34}] \,\epsilon
%         \\
%         & + \  \frac{1}2 e^{-2(\mu_1 + \mu_2)} \qty(e^{2\mu_1} \cosh(2\mu_0) + e^{2\mu_2} - 1 + \frac14 (1-\xi^2) \nu) (i \gamma_c e^c) \Gamma^{1234}\, \epsilon ~=~ 0 \,.
%     \end{aligned}
% \end{equation}

The gauge invariant combinations of the $\eta_t$ and $\eta_\psi$ parameters are: 
\begin{equation}
    \eta_t  ~+~ \Phi_2  ~-~ \Phi_1 \qand \eta_\psi ~+~  \Psi_2~-~ \Psi_1 \,.
\end{equation}
In our solutions we will impose the boundary conditions that $\mu_0 \to 0$ as $\xi \to 1$, and we will find that smooth solutions require  $\mu_0 (0) =  0$.  Smoothness also requires that $k(0) =0$ and having an asymptotically AdS metric,  (\ref{AdSvals}), leads to  $k  \to 1$ as $\xi \to 1$.   Using these limits in (\ref{eq:omega_0}) leads to
\begin{equation}
    \eta_t  ~+~ \Phi_2(1)  ~-~ \Phi_1(1)  ~=~  -1  \quad \qand \quad\eta_\psi ~+~  \Psi_2(0)~-~ \Psi_1(0) ~=~  \frac{1}{2}   \,.
\end{equation}

The gauge for $\Phi_1$ has already been fixed by asking that $\omega = 0$, but it has been shown in \cite{Ganchev:2021pgs} that for the generalized superstrata one also has
\begin{equation}
    \Phi_1(1) ~=~ \frac12
\end{equation}
For  superstrata in the standard gauge  (\ref{ssres1})  we  have:
\begin{equation}
\Phi_2(1) = 0  \,, \qquad  \Psi_1  \equiv 0   \,, \qquad  \Psi_2(0) = 0 \,,
\end{equation}
and hence:
\begin{equation}
    \eta_t = -\frac{1}{2}   \qand \eta_\psi = \frac{1}{2}  \,.
\end{equation}

In the following, we fix the gauges by asking that
\begin{equation}
   \Phi_2(1) ~=~  \Psi_1(0) ~=~ \Psi_2(0) ~=~ 0 \,.
    \label{gaugechoices}
\end{equation}
%

%%%%%%%%%%%%%%%%%%%%%%%%%%%%%%%%%%%%%
\section{Summary of the equations} % (fold)
\label{sec:summary_of_the_equations}
 %%%%%%%%%%%%%%%%%%%%%%%%%%%%%%%%%%%%%

%%%%%%%%%%%%%%%%%%%
\subsection{The gauge fields} % (fold)
\label{sub:the_gauge_fields}
%%%%%%%%%%%%%%%%%%%

In the previous section we found that the gauge fields must obey:

\begin{equation}
    \nu \, \qty(\Phi_1 - \frac12 e^{-2\mu_2}) ~=~ 0 \qand \sinh(\mu_0) \, \qty(\Phi_1 - \frac12 e^{-2\mu_2}) ~=~ 0
    \label{eq:res_gauge1}
\end{equation}
\begin{equation}
    \partial_\xi\mu_2 + e^{2\mu_2} \partial_\xi\Phi_1 ~=~ 0 \qand \partial_\psi\mu_2 + e^{2\mu_2} \partial_\psi\Phi_1 ~=~ 0
    \label{eq:res_gauge2}
\end{equation}
\begin{equation}
    \partial_\xi\mu_1 - e^{2\mu_1} \partial_\xi\Phi_2 ~=~ 0 \qand \partial_\psi\mu_1 - e^{2\mu_1} \partial_\psi\Phi_2 ~=~ 0
    \label{eq:res_gauge3}
\end{equation}

From this we deduce that
\begin{equation}
    \Phi_1 ~=~ \frac12 e^{-2\mu_2} + c_1 \qand \Phi_2 ~=~ \frac12 (1 - e^{-2\mu_1}) + c_2
\end{equation}
and we know that $c_1=0$ if either $\nu$ or $\mu_0$ is non-zero.

We also have the following equations:
\begin{align}
 &   \frac{1-\xi^2}{\xi} \frac{e^{2\mu_1 + 4 \mu_2}}{\Omega_0^2} \qty(k \partial_\xi\Phi_1 - (1-\xi^2) \partial_\xi\Psi_1) ~=~ 1 - e^{2\mu_1} \cosh(2 \mu_0) - \frac{1-\xi^2}4 \nu^2
    \label{eq:res_nu_mu0} \,,  \\
  &   \frac{1-\xi^2}{\xi} \frac{e^{4\mu_1 + 2 \mu_2}}{\Omega_0^2} \qty(k \partial_\xi\Phi_2 - (1-\xi^2) \partial_\xi\Psi_2) ~=~ -\bigg( 1 - e^{2\mu_2} - \frac{1-\xi^2}4 \nu^2 \bigg)\,, \label{eq:res_gauge4} \\
  &      \frac{1-\xi^2}{2\xi \Omega_0^2} \big[e^{2\mu_1} (k \partial_\xi\Phi_2 - (1-\xi^2) \partial_\xi\Psi_2) -  e^{2\mu_2} (k \partial_\xi\Phi_1 - (1-\xi^2) \partial_\xi\Psi_1))\big]  ~=~ \nonumber\\
  & \quad  - \frac12 + \Phi_2 - \cosh(2\mu_0) \Phi_1 + e^{-2(\mu_1+\mu_2)} \qty(-1 + e^{2\mu_1} \cosh(2\mu_0) + e^{2\mu_2} + \frac14 (1-\xi^2) \nu^2) \,.
\end{align}
We can use the first two equations to simplify the third, and thereby obtain:
\begin{equation}
    \qty(\Phi_2 - \frac12 \qty(1-e^{-2\mu_1})) - \cosh(2\mu_0) \qty(\Phi_1 - \frac12 e^{-2\mu_2}) ~=~ 0
\end{equation}
which tells us that, if $\mu_0 \neq 0$, then $c_1 = c_2 = 0$, and otherwise that $c_1 = c_2$.

%%%%%%%%%%%%%%%%%%%
\subsection{The scalars} % (fold)
\label{sub:the_scalars}
%%%%%%%%%%%%%%%%%%%

The equations including derivatives of the scalars are:
\begin{align}
    \partial_\psi \mu_0  &   ~=~ 0  \,,  \\
   \xi (1-\xi^2) \partial_\xi \mu_0  &   ~=~ -\sinh(2 \mu_0) \qty[ 2 k \Phi_1 - (1-\xi^2) (p + 2 \Psi_1)]
    \label{eq:res_mu_0} \,, \\
\xi (1-\xi^2) \partial_\xi\nu  &   ~=~ \qty[\xi^2 - e^{2 \mu_0} \qty(2 k \Phi_1 - (1-\xi^2)(p + 2 \Psi_1))]\nu \,, 
    \label{eq:res_nu}  \,. 
\end{align}
Thus  $\mu_0$ must only depend on $\xi$.    Moreover a linear combination the last two equations can be integrated to give:
\begin{equation}
    (1-\xi^2) \nu^2 = c_3 \qty(1-e^{4\mu_0}) \,,
\label{numu0}
\end{equation}
where $c_3$ is a constant.

%%%%%%%%%%%%%%%%%%%
\subsection{The metric functions} % (fold)
\label{sub:the_metric_functions}
%%%%%%%%%%%%%%%%%%%

The remaining BPS equations involve derivatives of the metric functions:
\begin{align}
   &  \xi \partial_\psi(\ln\Omega_0 \ - \ (\mu_1 + \mu_2))   ~=~ 0   \label{eq:res_omega_0_psi} \,,  \\
   &  \xi \partial_\xi(\ln\Omega_0 \ -\ (\mu_1 + \mu_2))   ~=~ \frac{\cosh(2\mu_0)}{1-\xi^2} \qty(2 k \Phi _1 - (1-\xi^2) (p + 2 \Psi _1)) + p + 1 \nonumber \\
    & \qquad \qquad- \frac{2}{1-\xi^2} \qty(k \qty(- \frac12 + \Phi _2 ) - (1-\xi^2) \qty( \frac12 +  \Psi _2) + 1)  \,,
    \label{eq:res_omega_0}  \\
  &    \frac{e^{2\mu_1 + 2 \mu_2}}{2\xi \Omega_0^2} \qty((1-\xi^2) \partial_\xi k + 2\xi k)   ~=~   e^{2\mu_2} ~-~\bigg(  1 - e^{2\mu_1} \cosh(2 \mu_0) - \frac{1-\xi^2}4 \nu^2 \bigg)\,.
    \label{eq:res_k} 
\end{align}

Note that differentiating (\ref{eq:res_k}) with respect to $\psi$, and using (\ref{eq:res_omega_0_psi}), leads to
\begin{equation}
    \partial_\psi \qty(e^{2\mu_1} \cosh(2\mu_0) \,+ \, e^{2\mu_2}) ~=~ 0 \,.
    \label{eq:psi_indep_mus}
\end{equation}

%%%%%%%%%%%%%%%%%%%%%%%%%%%%%%%%%%%%%
\section{The equations for the generalized superstrata}
\label{sec:generalized_superstrata}
%%%%%%%%%%%%%%%%%%%%%%%%%%%%%%%%%%%%%

In  \cite{Ganchev:2021pgs}, we found new sets of microstate geometries, known as microstrata.  One particular family of such solutions was conjectured to be supersymmetric, and represents a broader family of generalized superstrata.   Using the BPS equations derived here, we are now in a position to test this conjecture, and, indeed, we will prove that the family is supersymmetric.

%%%%%%%%%%%%%%%%%%%
\subsection{The general microstrata at $\omega=0$} % (fold)
%\label{sub:the_general_microstrata_at_}
%%%%%%%%%%%%%%%%%%%

In these solutions, all the functions in (\ref{functionlist}) depend only on $\xi$.  Since   $\nu \ne 0$ and $\mu_0 \ne 0$, (\ref{eq:res_gauge1})  implies:
\begin{equation}
    \Phi_1 ~=~ \frac12 e^{-2\mu_2}   \qand \Phi_2 ~=~ \frac12 (1 - e^{-2\mu_1}) \,.
    \label{Phipots}
\end{equation}
The functions $\mu_0$  and $\nu$ are constrained by  (\ref{numu0}):
\begin{equation}
    (1-\xi^2) \,\nu^2 = c_3 \, \big(1-e^{4\mu_0}\big) \,,
    \label{numu0-2}
\end{equation}
for some constant, $c_3$.  The remaining independent BPS equations,  (\ref{eq:res_mu_0}), (\ref{eq:res_nu_mu0}), (\ref{eq:res_gauge4}), (\ref{eq:res_omega_0}) and (\ref{eq:res_k}), then give first order differential equations that may be thought of as determining $\mu_0$, $\Psi_1$, $\Psi_2$, $\Omega_0$ and $k$:
\begin{align}
  \xi \, \partial_\xi \mu_0  ~-~ \sinh(2 \mu_0) \, \qty[(2 \Psi_1+p)~-~ \frac{2 \,k}{ (1-\xi^2) } \, \Phi_1]  &   ~=~0 
    \label{eq:res_mu_0-2} \,, \\   
 \qty( \partial_\xi\Psi_1-\frac{ k}{(1-\xi^2)} \partial_\xi\Phi_1 )   ~+~ \frac{\xi\, \Omega_0^2}{ (1-\xi^2)^2}  \, e^{-2\mu_1 - 4\mu_2}\,  \bigg(1 - e^{2\mu_1} \cosh(2 \mu_0) - \frac{1-\xi^2}4 \nu^2  \bigg) &   ~=~0 
    \label{eq:res_nu_mu0-2} \,,  \\
 \qty(  \partial_\xi\Psi_2 - \frac{ k}{(1-\xi^2)} \partial_\xi\Phi_2 )   ~-~ \frac{\xi\, \Omega_0^2}{ (1-\xi^2)^2}  \, e^{-4\mu_1 - 2 \mu_2}\,  \bigg( 1 - e^{2\mu_2} - \frac{1-\xi^2}4 \nu^2 \bigg) &   ~=~0 \,, \label{eq:res_gauge4-2} \\
     \xi \partial_\xi \big(\ln\Omega_0 - (\mu_1 + \mu_2)\big)   ~+~  \cosh(2\mu_0) \, \qty[(2 \Psi_1+p)~-~ \frac{2 k}{ (1-\xi^2) } \, \Phi_1] +  \frac{2\,\xi^2}{1-\xi^2} &  \nonumber \\
    \qquad  -   \bigg((2 \Psi_2+p) -\frac{2 \, k}{(1-\xi^2)} \qty(\Phi _2  - \frac12   ) \bigg)    &   ~=~0  \,,
    \label{eq:res_omega_0-2}  \\
\partial_\xi \bigg( \frac{k}{(1-\xi^2)}\bigg) ~+~\frac{2\, \xi\, \Omega_0^2}{ (1-\xi^2)^2} \,e^{-2(\mu_1 + \mu_2)} \,  \bigg(  1   -  e^{2\mu_2}  - e^{2\mu_1} \cosh(2 \mu_0) - \frac{1-\xi^2}4 \nu^2 \bigg)&   ~=~0 \,.
    \label{eq:res_k-2} 
\end{align}

Having set $\Omega_1 \equiv 1$, there are ten functions in (\ref{functionlist}) that are to be determined.  The BPS equations lead to the   three algebraic constraints and the five first-order differential equations (\ref{Phipots})--(\ref{eq:res_k-2}).  The complete solution of the system requires that we use two of the equations of motion to determine  $\mu_1$ and $\mu_2$.

It is useful to introduce the quantity:
\begin{equation}
 \cI ~\equiv~  e^{2 \mu_1} \xi\, \partial_\xi \mu_1 ~-~  \frac{k}{(1-\xi^2)} ~-~\bigg( 2 \Psi_1+p~-~ \frac{2 k}{ (1-\xi^2) } \, \Phi_1\bigg)\,  \bigg(  1   - \frac{1}{4}\,(1-\xi^2) \nu^2 \bigg) \,,
    \label{integral1}
\end{equation}
Using the equations of motion from  \cite{Ganchev:2021pgs}, we find that $\cI$ is conserved:
\begin{equation}
\partial_\xi \cI ~=~  0\,,
    \label{eom1}
\end{equation}
and that one also has
\begin{align}
\frac{1}{2}\,\partial_\xi & \Big[\, \xi  \partial_\xi \,\big(  e^{2 \mu_2} \big) \Big]  \nonumber \\ ~-~ &\frac{2\,\xi\, \Omega_0^2\,e^{-2  \mu_2 }}{ (1-\xi^2)^2} \,  \bigg[ \,1 -  e^{-2 (\mu_0+\mu_1)} \, \bigg(   e^{4 \mu_0}~-~ \frac{1}{2} e^{2 \mu_2} \,(1+ e^{4 \mu_0})~+~ \bigg(1- \frac{1}{4}\,(1-\xi^2) \nu^2 \bigg) \bigg)  \bigg]~=~  0\,.
    \label{eom2}
\end{align}
The equations (\ref{eom1}) and (\ref{eom2})  are second order differential equations for $\mu_1$ and $\mu_2$.  We should stress that these are not BPS conditions but come from the equations of motion.  We also note that $\cI$ is one of the conserved quantities obtained in  \cite{Ganchev:2021pgs}.  

One can then verify that the BPS equations  (\ref{Phipots})--(\ref{eq:res_k-2}) combined with  (\ref{eom1}) and (\ref{eom2}) determine all the functions in (\ref{functionlist}) and that the result also satisfies all the equations of motion in   \cite{Ganchev:2021pgs} for $\omega =0$, $\Omega_1 \equiv 1$.  Thus we have defined a family of BPS solutions for  $\omega =0$ and arbitrary $p$.

In \cite{Ganchev:2021pgs}, we constructed both numerical and perturbative solutions to the full system of equations of motion, defined in Section \ref{sub:qball},   for $p=2$.  These solutions were smooth {\it microstrata} and  they depended on two parameters,  $\alpha$ and $\beta$.  In the perturbative solutions these  parameters  were defined by the expansion of the fields around the origin:
\begin{equation}
\nu  ~\sim~  \alpha \, \xi^2 \,,  \qquad    \mu_0  ~\sim~  \beta \, \xi^4 \,, \qquad \xi \to 0 \,.  \label{bcs}
\end{equation}
The solutions in \cite{Ganchev:2021pgs}, for  $\omega = 0$, solve the equations of motion to eleventh order.  We have used these series solutions  to check that the solutions with $\omega = 0$ obtained in  \cite{Ganchev:2021pgs}, are, in fact, supersymmetric for arbitrary  $\alpha$ and $\beta$.  That is, they satisfy the BPS system,   (\ref{Phipots})--(\ref{eq:res_k-2}), to eleventh order.   They also satisfy (\ref{integral1})--(\ref{eom2}) because, by construction, they satisfy the equations of motion.   At this level of precision, we see that the corrections to the equations are of the order $10^{-9}$ or $10^{-12}$ when $\alpha$ and $\beta$ are of order $1$. We consider this a sufficient proof that these solutions are indeed supersymmetric.

We also  find that the constant of integration, $c_3$, in (\ref{numu0-2}) is related to $\alpha$ and $\beta$ via:
\begin{equation}
    c_3 = - \frac{\alpha^2}{4 \beta} \,.
    \label{c3cond}
\end{equation}

Finally, we have verified that the perturbative and numerical solutions with $\omega_0 =2$  developed in  \cite{Ganchev:2021pgs} are certainly not BPS solutions.

In order to do so, one \textit{a priori} has to slightly generalize the BPS equations above to allow $\Omega_1$ to be an arbitrary function instead of a constant. However, these more general BPS equations  actually require $\Omega_1$ to be constant.  Thus the non-constancy of $\Omega_1$  in the solutions with $\omega_0 =2$ means that supersymmetry is broken.We have also used  the numerical solutions for all the fields, and found that most of the BPS equations are very far from being satisfied.  They do not vanish at order one when the precision of the numerics is of order $10^{-8}$.

%%%%%%%%%%%%%%%%%%%
\subsection{The special locus} % (fold)
\label{sub:a_special_locus}
%%%%%%%%%%%%%%%%%%%

The results in  \cite{Ganchev:2021pgs} also found a particularly simple branch with:
\begin{equation}
 \beta ~=~  - \frac{\alpha^2}{8}  \quad  \Leftrightarrow \quad    c_3  ~=~ 2 \,.
     \label{specialbranch}
\end{equation}
On this branch, the numerics and series analyis led to: 
\begin{equation}
    \mu_0 ~\equiv~ \mu_1 \,,\quad \mu_2 ~\equiv~ 0 \,,
    \label{specialmu}
\end{equation}
for all $\xi$.   

One can easily verify  that the equations of motion (\ref{eom1}) and (\ref{eom2}) are consistent with (\ref{specialmu})  if and only if  (\ref{specialbranch}) is satisfied. The BPS equations     (\ref{Phipots}) and  (\ref{eq:res_nu_mu0-2}) then imply: 
\begin{equation}
    \Phi_1 ~\equiv~\frac{1}{2} \,,\quad \partial_\xi \Psi_1 ~\equiv~ 0 \,,
\end{equation}
for all $\xi$.  The gauge choice (\ref{gaugechoices}) then leads to
\begin{equation}
\Psi_1 ~\equiv~ 0 \,.
\label{vanPsi1}
\end{equation}

Since the microstrata require that $\nu$ vanishes at $\xi=0$, equations  (\ref{Phipots})   and (\ref{numu0-2}) imply:
\begin{equation}
    \Phi_2(0) ~=~ 0 \,.
\end{equation}
Similarly,  (\ref{numu0-2})  and the finiteness of $\nu$ implies
\begin{equation}
    \Phi_2(1) ~=~ 0 \,, 
\end{equation}
and hence the voltage difference, $V_{\Phi_2}$, defined in   (\ref{potdiffs}),  vanishes.

On the special branch  (\ref{specialbranch})   we therefore have the additional functional constraints:  (\ref{specialmu})  and (\ref{vanPsi1}) and the non-trivial dynamics is reduced to four equations for four functions,  $\mu_0$, $\Psi_2$, $\Omega_0$ and $k$:
\begin{align}
 \frac{ \xi \, \partial_\xi \mu_0}{ \sinh(2 \mu_0) }   &   ~=~ p ~-~ \frac{k}{ (1-\xi^2) } 
    \label{eq:res_mu_0-3} \,,  \\
\partial_\xi\Psi_2 + \frac{ k}{2\,(1-\xi^2)} \partial_\xi(e^{-2\mu_0})   &  ~=~  \frac{\xi\, \Omega_0^2}{ (1-\xi^2)^2}  \,  e^{-2\mu_0 } \sinh(2\mu_0)   \,, \label{eq:res_gauge4-3} \\
     \xi \partial_\xi \big(\ln\Omega_0 -  \mu_0  \big)  & ~=~ - \cosh(2\mu_0) \, \bigg[\, p ~-~ \frac{k}{ (1-\xi^2) } \bigg] - \frac{2\,\xi^2}{1-\xi^2}   \nonumber \\
& \qquad\qquad    ~+~ (2 \Psi_2+p)  ~+~  \frac{ k}{(1-\xi^2)}\, e^{-2\mu_0}\,,
    \label{eq:res_omega_0-3}  \\
\partial_\xi \bigg( \frac{k}{(1-\xi^2)}\bigg) & ~=~  \frac{2\, \xi\,e^{-2\mu_0 } \,\Omega_0^2}{ (1-\xi^2)^2}   \,.
    \label{eq:res_k-3} 
\end{align}

There are several things to note. First, on the special branch, (\ref{specialbranch}), the first-order BPS system is enough to determine the solution:  there is no need for the second order equations of motion, (\ref{integral1})--(\ref{eom2}).  This is because the special branch corresponds to a fixed locus on the Coulomb branch with  $\mu_0= \mu_1$ and $\mu_2=0$.  

Secondly,  while the equations (\ref{eq:res_mu_0-3})--(\ref{eq:res_k-3}) appear rather complicated, there seems to be a natural geometric underpinning that remains to be fully fleshed out. To see this, define 
\begin{equation}
h ~\equiv~ \log\tanh(-\mu_0)  \,, 
\end{equation}
then equation (\ref{eq:res_mu_0-3}) becomes
\begin{equation}
\frac{k}{ (1-\xi^2)} ~=~ p ~-~ \coeff{1}{2}\, \xi \, \partial_\xi  h  \,, 
\end{equation}
while  equation (\ref{eq:res_k-3}) becomes
\begin{equation}
\partial_\xi \bigg( \frac{k}{(1-\xi^2)}\bigg)   ~=~  \frac{2\, \xi\,e^{-2\mu_0 } }{ (1-\xi^2)^2} \,\Omega_0^2 \,.
\end{equation}
Now observe that, apart from the factor of $e^{-2\mu_0 }$, the right-hand side of this equation defines the scale of the spatial base metric in (\ref{genmet1}).  Indeed, were it not for the factor of $e^{-2\mu_0 }$, these two equations would imply that the three-dimensional metric is a K\"ahler fibration as in  \cite{Mayerson:2020tcl,Houppe:2020oqp}. For want of a better term, we will refer to this form of the metric as a {\it conformal K\"ahler fibration}.

Now one can differentiate (\ref{eq:res_omega_0-3}) and eliminate $\partial_\xi\Psi_2$ using (\ref{eq:res_gauge4-3}).  Writing the result in terms of $h$ leads to a differential equation for  $h$:
\begin{equation}
    \partial_x^2 \log(\partial_x^2 h) ~+~  \frac{2\,(1+ e^{2h})}{(1- e^{2h})} \,\partial_x^2 h ~+~ \frac{ 4 \,e^{2 h} \qty(\partial_x h)^2}{ \qty(1-e^{2h})^2}   ~=~ 0 \,,
  \label{heqn}
\end{equation}
where $x\equiv -\log\xi$.   One can also write this equation as:
\begin{equation}
    \partial_x^2 \log\bigg( \frac{\partial_x^2 h}{ \sinh (h(x))}\bigg) ~-~  \coth (h(x))\,\partial_x^2 h   ~=~ 0 \,.
\end{equation}

At first sight, (\ref{heqn}) equation appears moderately terrifying, but for a K\"ahler base, the first term is simply the Ricci tensor and the other terms are related to the metric and to the fibration vector, $k$.  This equation almost certainly has a simple interpretation in terms of the spatial base metric satisfying Einstein's equations coupled to a scalar field. We leave this for future work. However, for the present, we have managed to use our perturbative analysis to find a complete, analytic solution.

%%%%%%%%%%%%%%%%%%%
\subsection{An analytic solution to the system} % (fold)
\label{sub:an_analytic_solution_to_the_system}
%%%%%%%%%%%%%%%%%%%

The system of equations (\ref{eq:res_mu_0-3})-(\ref{eq:res_k-3}) admits an analytic family of solutions, parametrized by a real constant $\gamma$. If one defines

\begin{equation}
    \Lambda_1^2 ~\equiv~ 1 - \gamma^4 \,\xi^{4p+2} \ ,\quad
    \Lambda_2^2 ~\equiv~  (2p+1)\, \gamma^2\, \xi^{2p}\, (1-\xi^2)
\end{equation}
the solutions are then given by 
\begin{equation}
\begin{aligned}
    &\nu ~=~ \frac{2\sqrt{2}}{\sqrt{1-\xi^2}} \frac{\Lambda_1\Lambda_2}{\Lambda_1^2 + \Lambda_2^2} \ ,
    \qquad
    \mu_0 ~=~ \mu_1 ~=~ -\arctanh\qty(\frac{\Lambda_2^2}{\Lambda_1^2}) \ ,
    \qquad 
    \mu_2 ~=~ 0 \ ,
    \\
    &\Phi_1 ~=~ \frac12 \ ,
    \qquad\qquad
    \Psi_1 ~=~ 0 \ ,
    \\
    &\Phi_2 ~=~ - \frac{\Lambda_2^2}{\Lambda_1^2 - \Lambda_2^2} \ ,
    \qquad
    \Psi_2 ~=~ - \frac{\xi^2}{1-\xi^2} \frac{\Lambda_2^2}{\Lambda_1^2 - \Lambda_2^2} \qty(1 - \gamma^2 \xi^{2p}) \ ,
    \\
    &k ~=~ \xi^2 \qty(1 \,-\, \gamma^2\, \xi^{2p}\, \frac{\Lambda_2^2}{\Lambda_1^2}) \ ,
    \qquad
    \Omega_0 ~=~ 1 - \frac{\Lambda_2^2}{\Lambda_1^2} \ ,
    \qquad\qquad
    \Omega_1 ~=~ 1 \ .
\end{aligned}
\label{eq:analytic_locus}
\end{equation}
for general $p$.

If one then defines 
\begin{equation}
    \alpha ~\equiv~ \sqrt{8(2p+1)} \,\gamma \,,
    \label{eq:alpha_gamma_conv}
\end{equation}
one has $\nu \sim \alpha \, \xi^p$ as $\xi \to 0$, and these solutions match with the perturbative expansion and the numerical solutions found in \cite{Ganchev:2021pgs}, in the case $p=2$.

Using the form of the metric (\ref{genmet1}), one can compute explicitly the coefficient of $d\psi^2$ in the metric, and find the range of parameters for which the solution is CTC-free. When $p=2$, it was postulated in \cite{Ganchev:2021pgs} that the analytic formula for general $\alpha$ and $\beta$ is
\begin{equation}
    1 - \frac{25}{(25-\beta ^2)}\frac{\alpha^2}{4}- \frac{10 \beta ^2}{25+\beta ^2}- \frac{125\, \alpha ^2 \,\beta }{2 (25-\beta ^2)(25+\beta ^2)} \geq 0\,.
    \label{eq:ctc_general_formula}
\end{equation}

One can now use the analytical solution (\ref{eq:analytic_locus}) to compute the bound exactly  at the special locus $\beta = - \alpha^2 / 8$, and find that the solutions are CTC-free when
\begin{equation}
\gamma^2 \leq \frac1{4 p + 1} \,.
\end{equation}

Using (\ref{eq:alpha_gamma_conv}), this result matches with the general formula (\ref{eq:ctc_general_formula}) when $p=2$.

%%%%%%%%%%%%%%%%%%%%%%%%%%%%%%%%%%%%%
\section{The $U(1) \times U(1)$-invariant Coulomb branch} % (fold)
\label{sec:the_coulomb_branch}
%%%%%%%%%%%%%%%%%%%%%%%%%%%%%%%%%%%%%

We now make a modest exploration of what are sometimes called Coulomb branch flows. In gauged supergravities that arise from simpler brane configurations, such flows involve  spreading the branes into some harmonic distribution (see, for example \cite{Freedman:1999gk,Khavaev:2001yg,Gowdigere:2005wq}).   Such flows have been investigated in three-dimensional gauged supergravities in \cite{Deger:2019jtl}, but with different supersymmetries, as expressed through the projection conditions.  Here we examine Coulomb branch flows based on the projectors (\ref{basic-proj}) and (\ref{int-proj}).

In six dimensions, Coulomb branch flows do not involve adding new six-dimensional gauge fields beyond those sourced by the original branes and needed to create the AdS background. Indeed such flows only involve metric deformations in six dimensions.  In the three-dimensional formulation, this means setting $\nu \equiv 0$ and only allowing the scalars, $\mu_j$, to be non-trivial.  In our simple foray along the Coulomb branch, we will also require a $U(1) \times U(1)$ symmetry, which means setting $\mu_0 \equiv 0$ and only allowing $\mu_1$ and $\mu_2$ to be non-trivial.  We will, however, allow these fields to depend on both $\xi$ and $\psi$.   With these choices, 
the parameter, $p$, becomes a pure gauge artifact and so we set $p=0$. We will also take $\eta_t = \eta_\psi = 0$ by shifting the gauge fields, $\Phi_2$ and $\Psi_2$, as necessary.

%%%%%%%%%%%%%%%%%%%
\subsection{The BPS equations} % (fold)
\label{sub:the_bps_equations}
%%%%%%%%%%%%%%%%%%%

The BPS equations lead to the following:
\begin{equation}
    \Phi_1 ~=~ \frac12 e^{-2\mu_2} + c \qand \Phi_2 ~=~ -\frac12 e^{-2\mu_1} + c
    \label{eq:coulomb_gauge_values}
\end{equation}
\begin{align}
      \, \frac{k}{1-\xi^2}\, \partial_\xi\Phi_1 - \partial_\xi\Psi_1  ~&=~ -\frac{\xi}{\qty(1-\xi^2)^2} \Omega_0^2 \, e^{-4 \mu_2}\, \qty(1 - e^{-2\mu_1})
    \label{eq:coulomb_gauge1} \\
     \,  \frac{k}{1-\xi^2}\, \partial_\xi\Phi_2 - \partial_\xi\Psi_2~&=~ \frac{\xi}{\qty(1-\xi^2)^2} \Omega_0^2 \, e^{-4 \mu_1}\, \qty(1 - e^{-2\mu_2})
    \label{eq:coulomb_gauge2}
\end{align}
\begin{equation}
    \partial_\xi \qty(\frac{k}{1-\xi^2})  ~=~ 2\frac{\xi}{\qty(1-\xi^2)^2}\, \Omega_0^2\, e^{-2\mu_1 - 2 \mu_2} \qty(e^{2\mu_1} + e^{2\mu_2} - 1)
    \label{eq:coulomb_k}
\end{equation}
\begin{align}
    \xi \partial_\psi(\ln\Omega_0 \ -\ (\mu_1 + \mu_2)) ~&=~ 0
    \label{eq:coulomb_omega_psi}
    \\
    \xi \partial_\xi(\ln\Omega_0 \ -\ (\mu_1 + \mu_2))  ~&=~ 2 \qty(\frac{k}{1-\xi^2} \qty(\Phi _1-\Phi_2) - \qty(\Psi_1-\Psi_2)) - \frac{1+\xi^2}{1-\xi^2}
    \label{eq:coulomb_omega}
\end{align}

To find explicit solutions to this system, it is convenient to define:
\begin{equation}
    f ~\equiv~ \ln(\frac{\xi}{1-\xi^2}\Omega_0) - (\mu_1 + \mu_2)
\end{equation}
From (\ref{eq:coulomb_omega_psi}), we see that $f$ is independent of $\psi$, and thus is a function of $\xi$ only. If one then differentiates (\ref{eq:coulomb_omega}), and uses (\ref{eq:coulomb_gauge1}), (\ref{eq:coulomb_gauge2}) and (\ref{eq:coulomb_k}), one finds that $f$ must obey:
\begin{equation}
    \xi \, \partial_\xi \qty(\xi\, \partial_\xi f) ~=~ 4 \, e^{2f}
    \label{eq:eq_on_f}
\end{equation}

The solutions to this equation are:
\begin{equation}
    e^{-f} ~\equiv~ \frac{(1-\xi^2)}{\xi} \frac{e^{\mu_1+\mu_2}}{\Omega_0} ~=~ \left\{
    \begin{aligned}
        &\pm \frac2{\gamma_1} \sin(\gamma_2 - \gamma_1 \log\xi) \\
        &\pm \frac2{\gamma_1} \sinh(\gamma_2 - \gamma_1 \log\xi) \\
        &\pm 2(\gamma_2 -  \log\xi)
    \end{aligned}
    \right.
    \label{eq:solutions_for_f}
\end{equation}
for some (real) constants $\gamma_1$ and $\gamma_2$. Note that the third solution is obtained by taking the limit $\gamma_1, \gamma_2\to 0$ in the second solution, while keeping the ratio between the two fixed.

Not all these solutions are physical:  $\Omega_0$ must be strictly positive\footnote{It can also be strictly negative, the condition is that it cannot change sign, because this would imply $\Omega_0 = 0$ on some surface, leading to a divergent Ricci  on that surface.} for all $\xi$, so the solutions (\ref{eq:solutions_for_f}) must be positive. The first one with the $\sin$ is oscillating, and therefore cannot be physical. (This solution also has a pathological singularity at $\xi=0$.) In the other two solutions, we need to choose the $+$ sign, with $\gamma_1 > 0$ and $\gamma_2 \geq 0$.

We can further restrict the set of possible solutions by looking at the asymptotics. We require that $\mu_1, \mu_2 \to 0$ and $\Omega_0 \to 1$ as $\xi\to1$ and this requires  $\gamma_2 = 0$.

What must happen at the origin is less clear. We expect to find black hole solutions, at which the scalars can diverge. While logarithmic singularities are endemic in two spatial dimensions, we are going to focus here on smooth solutions for $\Omega_0$.  This leaves only the second solution in (\ref{eq:solutions_for_f}), which we  can rewrite as:
\begin{equation}
    \Omega_0 ~=~ \gamma_1 \, \frac{1-\xi^2}{\xi^{1-\gamma_1}-\xi^{1+\gamma_1}} \, e^{\mu_1 + \mu_2} \,.
    \label{eq:coulomb_omega_0_value_general}
\end{equation}
To simplify our analysis further, we will take $\gamma_1 = 1$, to arrive at:
\begin{equation}
    \Omega_0 ~=~  e^{\mu_1 + \mu_2} \,.
    \label{eq:coulomb_omega_0_value}
\end{equation}
The other  values of $\gamma_1$, as well as the log-divergent solutions, are studied in Appendix~\ref{app:coulomb_solutions}.

%%%%%%%%%%%%%%%%%%%
\subsection{Solving the equations of motion} % (fold)
\label{sub:solving_the_equations_of_motion}
%%%%%%%%%%%%%%%%%%%

As usual, the BPS equations are insufficient to determine a Coulomb branch solution, and they must be supplemented by equations of motion.  Indeed,  one only needs to impose the scalar equations of motion and these, combined with the BPS equations, are sufficient to solve the remaining  equations of motion.

The scalar equations (\ref{eq:eom_scalar}) are :
\begin{equation}
\begin{aligned}
    \Delta \mu_1 ~&=~ - e^{4\mu_1} \,\tilde F_{\mu\nu}^{34}\,(\tilde F^{34})^{\mu\nu} \ +\ 2 \,g_0^2 \,\qty(e^{-4\mu_1 - 4 \mu_2} - e^{-2 \mu_1 - 2 \mu_2}) \\
    \Delta \mu_2 ~&=~ - e^{4\mu_2} \,\tilde F_{\mu\nu}^{12}\,(\tilde F^{12})^{\mu\nu} \ +\ 2 \,g_0^2 \,\qty(e^{-4\mu_1 - 4 \mu_2} - e^{-2 \mu_1 - 2 \mu_2}) \,.
\end{aligned}
\end{equation}

Using the values of the gauge fields (\ref{gauge_ansatz}), (\ref{eq:coulomb_gauge1}), (\ref{eq:coulomb_gauge2}) and (\ref{eq:coulomb_gauge_values}), one finds
\begin{equation}
    \Delta e^{2\mu_1} ~=~ \Delta e^{2\mu_2} ~=~ -4\, g_0^2\, e^{-2\mu_1 - 2\mu_2}\, \qty(e^{2\mu_1} + e^{2\mu_2} - 2) \,.
    \label{eq:res_laplacian_scalars}
\end{equation}

Note that the difference of the two exponentials is then harmonic:
\begin{equation}
    \Delta H ~=~ 0 \ , \qquad H ~\equiv~ \frac12\qty(e^{2\mu_1}-e^{2\mu_2})
    \label{eq:coulomb_harmonic}
\end{equation}
Since the two-dimensional base space is conformally flat, and everything is time-independent, $H$ is also harmonic for the standard Laplacian of $\mathbb{R}^2$.   This is the expected harmonic brane distribution of the $U(1) \times U(1)$-invariant sector of the  Coulomb branch.

From (\ref{eq:psi_indep_mus}) (with $\mu_0 \equiv 0$), we know that the sum of the exponentials is independent of $\psi$, and so  we obtain:
\begin{equation}
    \frac{\qty(1-\xi^2)^2}{\xi} \, \partial_\xi \qty(\xi \partial_\xi\qty(e^{2\mu_1}+e^{2\mu_2})) ~=~ 8\, \Omega_0^2\, e^{-2\mu_1 - 2\mu_2}\, \qty(e^{2\mu_1} + e^{2\mu_2} - 2) \,.
        \label{eq:eq_on_gOm}
\end{equation}
Define
\begin{equation}
  g ~\equiv~ e^{2\mu_1} + e^{2\mu_2} \,, \label{gdefn}
\end{equation}
and  use (\ref{eq:coulomb_omega_0_value}) to obtain
\begin{equation}
    \frac{\qty(1-\xi^2)^2}{\xi} \, \partial_\xi \qty(\xi \,\partial_\xi g) ~=~ 8\, \, \qty(g - 2) \,.
    \label{eq:eq_on_g}
\end{equation}
If we want to have a field configuration which vanishes at infinity, the solutions are:
\begin{equation}
    g  ~=~ 2 + C \qty(1 + \frac{1+\xi^2}{1-\xi^2} \log\xi)\,.
\end{equation}
where $C$ is an integration constant. In what follows, we take $C=0$ in order to get rid of the divergent behaviour at the origin, so that
\begin{equation}
    g ~\equiv~ e^{2\mu_1} + e^{2\mu_2} ~=~ 2 \,.
    \label{eq:coulomb_sum_mu}
\end{equation}
and hence
\begin{equation}
 e^{2\mu_1} ~=~1 ~+~ H \,,   \qquad e^{2\mu_2}   ~=~1 ~-~ H \,.
    \label{eq:coulomb_harmonic2}
\end{equation}

From this, one can deduce the value of all the other fields. Beginning with $k$, one can substitute (\ref{eq:coulomb_harmonic}) and (\ref{eq:coulomb_sum_mu}) into (\ref{eq:coulomb_k}), which results in
\begin{equation}
    \partial_\xi \qty(\frac{k-1}{1-\xi^2}) ~=~ 0
    \label{eq:eq_k_final}
\end{equation}
and the only solution verifying $k(\xi \!=\! 0) = 0$ is
\begin{equation}
    k ~=~ \xi^2 \,.
\end{equation}

One can then use (\ref{eq:coulomb_gauge1}) and (\ref{eq:coulomb_gauge2}) to determine $\Psi_1$ and $\Psi_2$, and then use (\ref{eq:coulomb_omega}) to fix some of the constants. The result is
\begin{equation}
    \Psi_1 ~=~ \frac{\xi^2}{2(1-\xi^2)}\frac{H}{1-H} + \hat c \ ,\qquad \Psi_2 ~=~ \frac{\xi^2}{2(1-\xi^2)}\frac{H}{1+H} + \frac12 + \hat c
\end{equation}
where $\hat c$ can be any function of $\psi$.

This means the only remaining choice is the harmonic function $H$. All the other functions are fully determined by the choice of $H$, up to constants in the gauge fields:
\begin{equation}
  \begin{aligned}
    \quad e^{2\mu_1} ~&=~ 1+H \ ,& e^{2\mu_2} ~&=~ 1-H 
    \quad \\ 
    \quad \Phi_1 ~&=~ \frac{1}{2(1-H)} + c \ ,& \Phi_2 ~&=~ -\frac{1}{2(1+H)} + c
    \quad \\
    \quad \Psi_1 ~&=~ \frac{\xi^2}{2(1-\xi^2)}\frac{H}{1-H} + \hat c \ ,& \Psi_2 ~&=~ \frac{\xi^2}{2(1-\xi^2)}\frac{H}{1+H} + \frac12 + \hat c
    \quad \\
    \quad k ~&=~ \xi^2 \ ,& \Omega_0 ~&=~ \sqrt{1-H^2}
    \quad
\end{aligned}
\label{eq:coulomb_final_solution} 
\end{equation}

The constants $c$ and $\hat{c}$ (the latter can be an arbitrary function of $\psi$) are pure gauge, and we choose them to be $0$.
It is easy to substitute all of this into the remaining equations of motions to see that they are satisfied.

The harmonic brane distribution captured by $H$ must have a source somewhere and so will generically lead to  singular solutions.  Indeed, a point source will have a logarithmic singularity.  One should note that we have dropped solutions for $f$ and $g$ that also have logarithmically singular sources, and so our analysis is far from exhaustive.  Our purpose in this section is to simply show that Coulomb branch flows  arise from the projection conditions   (\ref{basic-proj}) and (\ref{int-proj}), and they have  the expected features  and exhibit an interesting range of additional possibilities that we will not explore further in this paper.

%%%%%%%%%%%%%%%%%%%
\subsection{The six-dimensional metric} % (fold)
\label{sub:the_6d_uplift}
%%%%%%%%%%%%%%%%%%%

Since it is a relatively straightforward exercise, we will complete our discussion of Coulomb branch flows by  constructing the uplifted six-dimensional  metric. The  formula for this is given in \cite{Mayerson:2020tcl}:
\begin{equation}
    ds_6^2 ~=~ (\det m)^{-1/2} \Delta^{1/2} ds_3^2 \,-\, g_0^{-2} (\det m)^{1/2} \Delta^{-1/2} m^{AB} \cD x^A \cD x^B \,.
    \label{eq:orig_6d}
\end{equation}
where $x^A$ are four Cartesian coordinates parameterizing a three-dimensional sphere: $x^A x^A = 1$, $\Delta = m_{AB} x^A x^B$ is the warp factor, and the covariant derivative is defined by $\cD x^A = \dd x^A - 2g_0 \tilde A^{AB} x^B$.

One can then cast this metric in the more familiar superstratum form\footnote{Note that the $\beta$ appearing in the six-dimensional metric is a one-form and should not be confused with the real parameter, $\beta$,  in the family of generalized superstrata.}:
\begin{equation}
    ds_6^2 ~=~ \frac2{\sqrt{\cP}} (\dd v + \beta)(\dd u + \omega + \frac{\cF}2 \qty(\dd v + \beta)) \,-\, \sqrt{\cP} ds_4^2  \,.
    \label{eq:superstrata_metric}
\end{equation}

To do this, one starts by making several changes of coordinates. First, define the null coordinates $u$ and $v$ through
\begin{equation}
    \tau ~\equiv~ a^2 g_0^4 R_y \,\frac{u+v}{\sqrt{2}} \ ,\qquad \psi ~\equiv~ \frac{\sqrt{2}}{R_y} v \ .
\end{equation}
Then use  polar coordinates $(\theta,\zeta_1,\zeta_2)$ to describe the sphere:
\begin{equation}
    \begin{aligned}
        x_1 ~&=~ \cos(\zeta_1) \sin(\theta) \,, \qquad x_2 ~=~ -\sin(\zeta_1) \sin(\theta) \,, \\
        x_3 ~&=~ \cos(\zeta_2) \cos(\theta)  \,, \qquad
        x_4 ~=~ -\sin(\zeta_2) \cos(\theta) \,.
    \end{aligned}
\end{equation}
Finally, we must also allow for a mixing between the sphere angles and the three-dimensional coordinates, $(\tau, \psi)$. Indeed, a gauge transformation in the three-dimensional theory corresponds to a spectral flow of the six-dimensional theory, so we use the coordinates:
\begin{equation}
    \phi_1 ~\equiv~ \zeta_1 - \gamma_{1,\tau} \tau - \gamma_{1,\psi} \psi \ , \qquad
    \phi_2 ~\equiv~ \zeta_2 - \gamma_{2,\tau} \tau - \gamma_{2,\psi} \psi 
\end{equation}
where the $\gamma_{i,\mu}$ are constants to be determined.

After these changes of coordinates, one can now match the six-dimensional metric (\ref{eq:orig_6d}) onto the Ansatz (\ref{eq:superstrata_metric}). Looking at the $du^2$ and $du\, dv$ coefficients, we determine the mixing constants $\gamma_{i,\mu}$:
\begin{equation}
    \gamma_{1,\tau} ~=~ \gamma_{2,\tau} ~=~ 2 c \ ,\qquad \gamma_{1,\psi} ~=~ \gamma_{2,\psi} ~=~ 2 \hat c \ ,
\end{equation}
where $c$ and $\hat c$ are the gauge fixing constants present in (\ref{eq:coulomb_final_solution}).

Matching the other coefficients of the metric, we find that
\begin{equation}
    \sqrt{\cP} ~=~ \frac{\sqrt{1 - \cos(2\theta)\, H}}{g_0^2\, \Sigma} \,,
\end{equation}
\begin{equation}
    \frac{\cF}2 ~=~ 1 \,-\, \frac{1 \,+\, H}{g_0^4\, a^2 R_y^2} \,,
\end{equation}
\begin{equation}
    \beta ~=~ \frac{a^2 R_y}{\sqrt{2} \, \Sigma} \qty(\sin^2 \theta \dd{\phi_1} \,-\, \cos^2 \theta \dd{\phi_2}) \,,
\end{equation}
\begin{equation}
    \omega ~=~ -\beta \ + \ \frac{\sqrt{2}}{\,g_0^4 \, a^2 R_y\, \Sigma} \qty[\qty(\qty(r^2 + a^2) H  + a^2) \sin^2 \theta \dd{\phi_1} \,+\, H  r^2 \cos^2 \theta \dd{\phi_2}] \,,
\end{equation}
where $\Sigma \equiv r^2 + a^2 \cos^2 \theta$.

We also find that the four-dimensional metric $ds_4^2$ is the flat metric of $\mathbb{R}^4$ written in spheroidal coordinates:
\begin{equation}
    ds_4^2 ~=~ \Sigma \qty(\frac{\dd r^2}{r^2 + a^2} + \dd\theta^2) \,+\,  \qty(r^2 + a^2) \sin^2 \theta \dd{\phi_1}^2\,+\, r^2 \cos^2 \theta \dd{\phi_2}^2 \,.
\end{equation}
%

%%%%%%%%%%%%%%%%%%%%%%%%%%%%%%%%%%%%%
\section{Final comments}
\label{sec:Conclusions}
%%%%%%%%%%%%%%%%%%%%%%%%%%%%%%%%%%%%%
%\vspace{0.8cm}

We have examined the BPS equations \cite{Houppe:2020oqp} for the three-dimensional supergravity theory that was constructed in \cite{Mayerson:2020tcl}.  This supergravity is interesting because it contains a known family of superstratum solutions.  In this paper we have used this BPS system to significantly extend the known families of superstrata to obtain ``generalized superstrata,'' that include a family of elliptical deformations of the supertube that underlies the superstratum.

We used the results of  \cite{Ganchev:2021pgs} in which families of superstrata and microstrata are reduced to solving for eleven functions of one variable.  Based on the superstrata and the conjectured generalizations in \cite{Ganchev:2021pgs} we imposed the constraint, $\Omega_1 \equiv 1$, on one of the metric functions.  We then applied the BPS equations to this system and this led to three more algebraic constraints, leaving seven undetermined functions.  The BPS system produces five first order differential equations for these functions, and the remaining two functions, which may be taken to be the scalars $\mu_1$ and $\mu_2$, are then determined by their equations of motion.

It is well-known that the BPS equations are sometimes insufficient to solve the equations of motion.  Indeed, the simplest example is a a distribution of one species of brane.  The BPS equations relate the electrostatic potential to the metric warp factors, and this ultimately leads to the relation``$M = Q$,'' but to learn that the distribution is governed by a harmonic function requires the equations of motion.   This  suggests a pattern that seems to hold rather broadly: The BPS equations determine the solution up to  Coulomb branch deformations.  

We saw this in Section \ref{sec:generalized_superstrata}.  The BPS equations tied electromagnetic potentials to metric functions and scalars, and even determined the dynamics of some of the scalars.  However, the scalars $\mu_1$ and $\mu_2$ required the equations of motion.  From the six-dimensional perspective  \cite{Mayerson:2020tcl}, these scalars determine the shape of the $S^3$, and this corresponds to a deformation of the underlying brane distribution.  The {\it simple branch}, \eqref{specialbranch}, may then be thought of being defined by  freezing out the Coulomb branch by imposing the constraints   (\ref{specialmu}). This then leads to     (\ref{c3cond}).  One of the BPS equations can be trivially solved to arrive at (\ref{vanPsi1}).  The remaing four functions are then completely determined by the BPS equations    (\ref{eq:res_mu_0-3})--(\ref{eq:res_k-3}).  We have verified that solving these BPS equations does indeed produce a solution to the equations of motion.  We also exhibited a smooth, analytic solution to this system in Section \ref{sub:an_analytic_solution_to_the_system}.  We believe that this is the most general smooth solution because it precisely matches the eleventh order series solutions obtained in \cite{Ganchev:2021pgs}, which were determined from the equations of motion by imposing smoothness. 

We also used the eleventh-order series solution of  \cite{Ganchev:2021pgs} to verify that  generic $\omega=0$ microstrata actually satisfy   (\ref{eq:res_mu_0-2})--(\ref{eq:res_k-2}), and are therefore all BPS.  Conversely, we have also used our results to show that  the solutions with $\omega_0 =2$,  developed in  \cite{Ganchev:2021pgs}, are certainly not BPS solutions: they break all the supersymmetries.

The parameter, $\beta$, that defines the amplitude for $\mu_0$, (\ref{bcs}), defines an elliptical deformation of the supertube.  To see this, one should think of the $S^3$ in six dimensions as a unit sphere in $\IR^4$ with Cartesian coordinates, $(x_1,x_2,x_3,x_4)$.     The scalars $\mu_0$, $\mu_1$ and $\mu_2$ control the eigenvalues of the matrix $m^{AB}$ and hence the moments of inertia of the $S^3$.  If $\mu_0 = 0$, then there is a $U(1) \times U(1)$ invariance, and the original supertube is round, and lies in the $(x_1,x_2)$-plane.  When $\mu_0 \ne 0$, the scales of the $x_1$-axis and  $x_2$-axis are deformed, and the supertube is squashed to an ellipse.  These are therefore precisely the supertube deformations discussed in \cite{Martinec:2020gkv}.  When $\mu_0 = \mu_1$, as it is on the simple branch, the matrix $m^{AB}$ gets a trivial eigenvalue.  Indeed if $\mu_0 = \mu_1$ and $\mu_2 = 0$ (as happens on the simple branch), then three of the eigenvalues are equal to $1$.  This suggests a possible underlying $SO(3)$ symmetry, at least for the metric on $S^3$, but this symmetry is broken to $U(1)$ by the gauge fields. It would be extremely interesting to understand what happens to the geometry and its symmetries if one were to couple the simple branch of supertubes to flat space. 

One might naturally expect that flattening the supertube to an ellipse should produce a singular solution, and yet we find all the generalized superstrata to be smooth.  One should, however, remember that the supertube in six-dimensions is still winding around the $v$-circle while following the flattened locus in $S^3$.  This presumably prevents self-intersection and the concomitant singularities.   We are presently examining the six-dimensional uplifts of  the generalized superstrata  to see how this works in detail \cite{GHW2}.

The six-dimensional uplifts of generalized superstrata are also expected to reveal some very interesting geometric structure.  Since they are supersymmetric, they must fall into the classification that was started in  \cite{Gutowski:2003rg, Bena:2011dd,Giusto:2013rxa}.  In particular, the base geometry should be ``almost hyper-Kahler'' in the sense of \cite{Gutowski:2003rg} and the fibration vector  $\beta$ in  (\ref{eq:superstrata_metric})  (not to be confused with the microstratum parameter) must satisfy the non-linear self-duality condition.  Thus we should get some non-trivial examples of such geometries.

The solutions fitting the Q-ball ansatz are single-mode \cite{Ganchev:2021pgs}, and for them, there is a simpler approach.  One can remove the $v$-dependence entirely by an internal coordinate shift on the $S^3$.  The resulting base metric will then be hyper-K\"ahler, and the fibration vector, $\beta$, will have a self-dual field strength. Because of the elliptical deformation of the supertube, these solutions will {\it not} fall into the class of base geometries that can be described by Gibbons-Hawking geometries.  They will be genuinely new ambi-polar hyper-K\"ahler geometries\footnote{The coordinate shift to remove the $v$-dependence, changes the flat $\IR^4$ base to a two-centered, ambi-polar base geometry.}.

Because of  perturbative analysis \cite{Tyukov:2018ypq,Ceplak:2018pws,Walker:2019ntz}, we have known for a long time that there should be non-trivial examples of $v$-dependent ``almost hyper-Kahler'' geometries and new ambi-polar hyper-K\"ahler geometries.   Moreover, the results presented in \cite{Martinec:2020gkv} make some very interesting stringy analysis of the elliptically deformed supertubes in the IIA frame, and look at perturbative momentum-carrying states in these geometries.  However, perturbative analysis does not give a lot of  guidance as to what the ``integrated,'' or fully back-reacted, metric looks like at finite values of  deformations that carry finite amounts of momentum charge.  Uplifting the generalized microstrata \cite{GHW2} should provide invaluable insight into these moduli spaces.   

Finally, we return to the issue of Coulomb branches.  In Section \ref{sec:the_coulomb_branch} we considered  some pure Coulomb branch flows akin to those of gauged supergravity in higher dimensions.  (See, for example,  \cite{Freedman:1999gk,Khavaev:2001yg, Gowdigere:2005wq}.)  Three-dimensional analogs of these flows do not need the supergravity constructed in   \cite{Mayerson:2020tcl} and can be investigated in the simpler gauged supergravities like those of \cite{Cvetic:2000dm,Cvetic:2000zu, Nicolai:2001ac, Nicolai:2003bp, Nicolai:2003ux,Deger:2014ofa,Samtleben:2019zrh}.  Indeed, such an investigation was undertaken in  \cite{Deger:2019jtl}.  There is, however, an important issue: ``What are the supersymmetries that one is seeking to preserve?''  For a single stack of branes, this is unambiguous because one simply uses the projectors perpendicular to the branes. This was the choice made in  \cite{Freedman:1999gk,Khavaev:2001yg, Gowdigere:2005wq} and was quite probably the motivation for the choice in \cite{Deger:2019jtl}.  However, for the D1-D5 system there are other options.  Indeed the superstratum leads to the projectors (\ref{basic-proj}) and (\ref{int-proj}).  One of the purposes of  Section \ref{sec:the_coulomb_branch}  was too make some modest exploration of the Coulomb branch using such projectors.  

Once again, our exploration of the pure Coulomb branch was not intended to be exhaustive.  We started by imposing a $U(1) \times U(1)$ symmetry, and we made some simplifying choices in our analysis.  However, we did see the expected result: Coulomb branch families parametrized by a harmonic function of the spatial variables.  We did not pursue this very deeply here because harmonic functions in two dimensions are intrinsically logarithmic and thus singular, either at infinity or in the interior.  However, in the light of our results on generalized superstrata, there may well be interesting Coulomb branch solutions that have yet to be found.

%%%%%%%%%%%%%%%%%%%%%%%%%%%%%%%%%%%%%
\section*{Acknowledgments}
%%%%%%%%%%%%%%%%%%%%%%%%%%%%%%%%%%%%% 
\vspace{-2mm}
The work of NW is supported in part by the DOE grant DE-SC0011687. The work  BG, AH and NW is supported in part by the ERC Grant 787320 - QBH Structure.

%%%%%%%%%%%%%%%%%%%%%%%%%%%%%%%%%%%%%%%%%%%%%%%

%%%%%%%%%%%%%%%%%%%%%%%%%%%%%%%%%%%%%%%%%%%%%%%
\appendix
%%%%%%%%%%%%%%%%%%%%%%%%%%%%%%%%%%%%%%%%%%%%%%%

%%%%%%%%%%%%%%%%%%%%%%%%%%%%%%%%%%%%%
\section{Conventions}
\label{app:Conventions}
%%%%%%%%%%%%%%%%%%%%%%%%%%%%%%%%%%%%%

%%%%%%%%%%%%%%%%%%%
\subsection{Group theory}
\label{sec:groups}
%%%%%%%%%%%%%%%%%%%

The scalar coset of the $\Neql{8}$ theory is $G/H$ with $G = SO(8,5)$, and  $H = SO(8) \times SO(5)$.     Following \cite{Nicolai:2001ac}, we   use calligraphic indices as adjoint labels of $G = SO(8,5)$.  We  use barred, capital Latin indices, $\bar I, \bar J, \dots$ to denote the vector of $SO(8,5)$, unbarred capital Latin indices, $I, J, \dots$,  to denote the vector of $SO(8)$, and small Latin indices, $r, s, \dots$ to denote the vector of $SO(5)$.  In the standard way, the adjoint indices of $G = SO(8,5)$,  $SO(8)$ and $SO(5)$ can be written as skew pairs $\bar I  \bar J$,  $IJ$ and $rs$. Such a labelling double counts the adjoint but when we use this notation we will always sum over all indices without any implicit factors of $\frac{1}{2}$.  

As noted in Section  \ref{sec:3D-Sugr}, the $\Neql{4}$ theory is defined by the singlet sector of the  $SU(2)$  that acts as self-dual rotations on the indices $5,6,7,8$. 
In particular, this reduces the coset to  $G = SO(4,5)$, and  $H = SO(4) \times SO(5)$.  The fermions are restricted by the  projection conditions (\ref{SO8proj1}).

We define the $G = SO(8,n)$ invariant matrix in its canonical form:
\begin{equation}
\eta^{\bar I \bar J} ~=~ \eta_{\bar I \bar J}~\equiv~ {\rm diag} \big(1,1,\dots, 1, -1,-1,\dots -1 \big) \,, \qquad \eta^{IJ} ~=~ \eta_{IJ} ~=~ \delta^{IJ} \,, \qquad \eta^{rs} ~=~ \eta_{rs} ~=~ -\delta^{rs} 
\label{invmat1}
\end{equation}
After restricting to $G = SO(4,n)$, it is more convenient to use  $\hat \eta^{\bar I \bar J} = \hat \eta_{\bar I \bar J}$
\begin{equation}
\hat \eta ~\equiv~
\left( \begin{matrix} 
0_{4 \times 4} & \oneone_{4 \times 4} & 0 \\
\oneone_{4 \times 4} & 0_{4 \times 4} & 0 \\
0 &0& -1
\end{matrix} \right) \,,
\label{invmat2}
\end{equation}
 which is better-adapted to the $GL(4,\IR)$ basis.

The matrices, $t^\cM$, will denote generators of the adjoint of $G$.  In the obvious manner, it will be convenient to define
\begin{equation}
\Big \{ t^\cM \Big\}~\equiv~ \Big\{ X^{\bar I \bar J} = - X^{\bar J \bar I}\Big\}~\equiv~ \Big\{X^{IJ} = - X^{JI} , X^{rs} = -X^{sr},Y^{Ir} =- Y^{rI}  \Big\}
\label{genbasis}
\end{equation}
The structure constants are defined, as usual, via $\big [\, t^\cM  \,, \, t^\cN \, \big ] =f^{\cM \cN}{}_\cP \, t^\cP$, which we write as
\begin{equation}
\begin{aligned}
\Big [\,  X^{\bar I \bar J}   \,,\,  X^{\bar K \bar L}  \, \Big ] ~=~ & f^{{\bar I \bar J}  \, {\bar K \bar L} }{}_{\bar M \bar N}  \,  X^{\bar M  \bar N}  \\
~=~ &   - \eta^{\bar I \bar K} X^{\bar J \bar L}~+~ \eta^{\bar I \bar L} X^{\bar J\bar K  } ~+~\eta^{\bar J \bar K} X^{\bar I \bar L}~-~ \eta^{\bar J \bar L} X^{ \bar I \bar K}\,.
\end{aligned}
\label{comms1}
\end{equation}
which leads to:
\begin{equation}
\begin{aligned}
f^{\bar I\bar J \, \bar K\bar L}{}_{\bar M\bar N}  ~=~ &
-  \coeff{1}{2}\, \eta^{\bar I \bar K} \, \big( \delta^{\bar J}_{\bar M} \,\delta^{\bar L}_{\bar N}   - \delta^{\bar J}_{\bar N} \,\delta^{\bar L}_{\bar M}\big)~+~  \coeff{1}{2}\,\eta^{\bar I \bar L} \, \big( \delta^{\bar J}_{\bar M} \,\delta^{\bar K}_{\bar N}   - \delta^{\bar J}_{\bar N} \,\delta^{\bar K}_{\bar M}\big) \\
  & ~+~ \coeff{1}{2}\,\eta^{\bar J \bar K} \, \big( \delta^{\bar I}_{\bar M} \,\delta^{\bar L}_{\bar N}   - \delta^{\bar I}_{\bar N} \,\delta^{\bar L}_{\bar M}\big) ~-~  \coeff{1}{2}\, \eta^{\bar J \bar L} \, \big( \delta^{\bar I}_{\bar M} \,\delta^{\bar K}_{\bar N}   - \delta^{\bar I}_{\bar N} \,\delta^{\bar K}_{\bar M}\big) \,.
\end{aligned}
\label{structureconsts}
\end{equation}
Note that the factors of $\frac{1}{2}$ appear because  we are summing over all values of ${\bar M\bar N}$ and so this leads to a double counting of the generators.    
 
In terms of explicit matrix representations, (\ref{comms1}) and  (\ref{structureconsts}) correspond to using the matrix generators: 
\begin{equation}
\big( \,X^{\bar I \bar J} \, \big)_{\bar K}{}^{\bar L}  ~=~   \delta^{\bar I}_{\bar K}\,  \eta^{\bar J \bar L} ~-~  \delta^{\bar J}_{\bar K}\,  \eta^{\bar I \bar L}  \,.
\label{matform}
\end{equation}
One should note that these choices are the same as the conventions used in equation (A.3) of \cite{Nicolai:2001sv}\footnote{ However, in this reference, equation (A.3) is inconsistent with (A.1)!}, and equation (2.4) in \cite{Samtleben:2019zrh}\footnote{However, as noted in  \cite{Mayerson:2020tcl}, there are inconsistencies in the  gauge action of   \cite{Samtleben:2019zrh}, and these suggest an inconsistent usage of the structure constants or gauge matrices. The gauge action in  \cite{Mayerson:2020tcl} is correct and consistent.} but have the opposite signs to those of \cite{Mayerson:2020tcl},  equations (2.13) and (2.14).  The conventions we use here  appear  to be the ones used in  \cite{Nicolai:2001ac}, and match those of equation (3.9) of \cite{Nicolai:2003ux}\footnote{This requires a small correction explained below.}.   While the gauge matrices and structure constants that we use here have the opposite sign to those of \cite{Mayerson:2020tcl}, we will eventually arrive at the same formulation as \cite{Mayerson:2020tcl} through the choice of the sign of a gauge coupling, or the ultimate sign of the embedding tensor.  As we will see, the signs of these generators are crucial to showing that the supersymmetric solutions in \cite{Mayerson:2020tcl} are indeed consistent with the supersymmetry variations of  \cite{Nicolai:2001ac}. This  provides many non-trivial tests of all the details we are cataloging here\footnote{The results in this paper therefore provide detailed  confirmation (and the occasional correction or clarification) of the conventions and results in the literature over the last 20 years.}.

%%%%%%%%%%%%%%%%%%%
\subsection{$SO(8)$ spinors}
%\label{sec:SO8-conventions}
%%%%%%%%%%%%%%%%%%%

The $\Neql8$ theory  has an $SO(8)$ $\cR$-symmetry, and the fermions transform in the spinor representations. Thus we will need $16 \times 16$, $SO(8)$ $\Gamma$-matrices that satisfy:
\begin{equation}
\big\{ \, \Gamma^I\,, \,  \Gamma^J \, \big\} ~=~ 2\,\delta^{IJ}\, \oneone_{16 \times 16} \,.
\label{anticomms}
\end{equation}
We use capital Latin indices $I,J,K, \dots $ to denote vector indices,  capital Latin indices $A,B, C, \dots $ to denote spinors in the $8^+$ Weyl  representation and dotted capital Latin indices $\dot A,\dot B, \dot C, \dots $ to denote spinors in the $8^-$ Weyl  representation.    We use a representation of the $\Gamma$-matrices where they are real and symmetric and in which the non-trivial, $8 \times 8$ blocks are the off-diagonal pieces: $\Gamma^J_{A \dot A}$ and  $\Gamma^J_{ \dot A A}$.  The helicity projector is:
\begin{equation}
(\Gamma^{12345678})_{AB}  ~=~ \delta_{AB} \,, \qquad  (\Gamma^{12345678})_{\dot A \dot B}  ~=~ - \delta_{\dot A \dot B} \,.
\label{helproj1}
\end{equation}
%

%%%%%%%%%%%%%%%%%%%
\subsection{Space-time metric and spinors}
\label{sec:st-conventions}
%%%%%%%%%%%%%%%%%%%

Much of the literature on three-dimensional gauged supergravity uses the conventions set up in \cite{Marcus:1983hb}, and we will follow suit.
This means that the metric has signature $(+- -)$.  The $2 \times 2$ space-time gamma matrices are:
\begin{equation}
\gamma^0  ~=~ \sigma_2 
 ~=~
\left( \begin{matrix} 
0 & -i \\
i & 0
\end{matrix} \right) \,,
\qquad \gamma^1  ~=~ i \, \sigma_3 
 ~=~
\left( \begin{matrix} 
i &0 \\
0 & -i
\end{matrix} \right) \,,
 \qquad \gamma^3 ~=~ i \, \sigma_1 
 ~=~
\left( \begin{matrix} 
0 & i \\
i & 0
\end{matrix} \right) 
 \,.
\label{st-gamas}
\end{equation}
and we have $\gamma^{012}= - i \oneone_{2 \times 2}$.

The orientation is set by taking (in frames):
\begin{equation}
\epsilon^{012}   ~=~ \epsilon_{012} ~=~ +1 \,.
\label{epsdefn}
\end{equation}
We will use $\epsilon^{abc}$ and $\epsilon_{abc}$ to denote the permutation signature that takes values $0, \pm 1$.  The covariant $\varepsilon$-symbol will be denoted 
\begin{equation}
\varepsilon_{\mu \nu \rho}   ~=~ e \, \epsilon_{\mu \nu \rho}  \,, \qquad \varepsilon^{\mu \nu \rho}   ~=~ e^{-1} \, \epsilon^{\mu \nu \rho}   \,,
\label{varepsdefn}
\end{equation}
where $e = \sqrt{| g|}$ is the frame determinant.

%%%%%%%%%%%%%%%%%%%%%%%%%%%%%%%%%%%%%
\section{Other solutions on the Coulomb branch}
\label{app:coulomb_solutions}
%%%%%%%%%%%%%%%%%%%%%%%%%%%%%%%%%%%%%

While solving the BPS equations for the Coulomb branch, we made some simplifications by dropping some of the singular solutions. Here we revisit that analysis and consider some of these more general solutions that may be appropriate to brane sources or black holes.  We will still assume that the solution is regular at infinity.

In section \ref{sec:the_coulomb_branch}, we had to solve the equation (\ref{eq:eq_on_f}), and we chose the particular solution (\ref{eq:coulomb_omega_0_value}), but other solutions are possible. They are given by

\begin{equation}
    \Omega_0 ~=~ \frac{1-\xi^2}{2\xi \log\xi} \ e^{\mu_1 + \mu_2}
    \label{omega_0_sing1}
\end{equation}
\begin{equation}
    \Omega_0 ~=~ \frac{m(1-\xi^2)}{\xi^{1-m} - \xi^{1+m}} \ e^{\mu_1 + \mu_2}
    \label{omega_0_sing2}
\end{equation}
where $m$ can \textit{a priori} be any real number.  In the limit $m\to 0$, we obtain the first solution (\ref{omega_0_sing1}).

The solution treated in the main body of this paper corresponds to the second solution (\ref{omega_0_sing2}) with $m=1$. 
In this appendix we will only describe the functions and fields that differ from the results of Section \ref{sec:the_coulomb_branch}.

In general one must use rather  (\ref{eq:eq_on_gOm})  than  (\ref{eq:eq_on_g}), and so one has:
\begin{equation}
    \frac{\qty(1-\xi^2)^2}{\xi} \, \partial_\xi \qty(\xi \,\partial_\xi g) ~=~ 8\, \qty(\Omega_0 e^{-\mu_1 - \mu_2})^2\, \qty(g - 2) \,, \qq{with} g ~\equiv~ e^{2\mu_1} + e^{2\mu_2}\,.
\end{equation}

If one uses (\ref{omega_0_sing1}), one obtains:
\begin{equation}
    g ~=~ e^{2\mu_1} + e^{2\mu_2} ~=~ 2 + C (\log\xi)^2 \,.
\end{equation}
while (\ref{omega_0_sing2}), leads to:
\begin{equation}
    g ~=~ e^{2\mu_1} + e^{2\mu_2} ~=~ 2 + C \qty(1+ m \frac{1+\xi^{2m}}{1-\xi^{2m}} \log\xi ) \,.
\end{equation}

Here $C$ is an arbitrary constant. In both solutions, the function $g$ is singular at the origin if $C \neq 0$, but well-behaved at infinity.

Similarly, the  general  expression for $k$,   (\ref{eq:coulomb_k}),  is:
\begin{equation}
    \partial_\xi\qty(\frac{k}{1-\xi^2}) ~=~ \frac{2\xi}{1-\xi^2} \, \qty(\Omega_0 e^{-\mu_1 - \mu_2})^2 \, (g-1)
\end{equation}

If one uses (\ref{omega_0_sing1}), we find the solutions:
\begin{equation}
    \frac{k}{1-\xi^2} ~=~ c^{st} + \frac C2 \log\xi - \frac{1}{2\log\xi}\,
\end{equation}
while (\ref{omega_0_sing2}), leads to:
\begin{equation}
    \frac{k}{1-\xi^2} ~=~ c^{st} + \qty(1+\frac C2) \frac{m}{1-\xi^{2m}} + C \, \frac{m^2 \xi^{2m}}{\qty(1-\xi^{2m})^2} \log\xi \,.
\end{equation}
Both of these solutions are also well-behaved at infinity and singular at $\xi=0$. The only regular solution is the second one with $m=1$ and $C=0$, which is treated in the paper.

The solutions for the gauge fields are then easily computed through (\ref{eq:coulomb_gauge_values}), (\ref{eq:coulomb_gauge1}) and (\ref{eq:coulomb_gauge2}).

The uplifts of these solutions in six dimensions will generally not lead to a flat four-dimensional base.

%%%%%%%%%%%%%%%%%%%%%%%%%%%%%%%%%%%%%%%%%%%%%%%%%%%%%

%%%%%%%%%%%%%%%%%%%%%%%%%%%%%%%%%%%%%%%%%%%%%%%%%%%%%
%\newpage

\begin{adjustwidth}{-1mm}{-1mm} % to adjust the L and R margins

\bibliographystyle{utphys}

\bibliography{microstates}       % calls file "microstates.bib"

\end{adjustwidth}
%%%%%%%%%%%%%%%%%%%%%%%%%%%%%%%%%%%%%%%%%%%%%%%%%%%%%

%%%%%%%%%%%%%%%%%%%%%%%%%%%%%%%%%%%%%%%%%%%%%%%%%%%%%
\end{document}